\documentclass[aps,twocolumn,showpacs,floatfix]{revtex4-1}

\usepackage{amsmath}
\usepackage{epsfig}
\usepackage{epsf}
\usepackage{array}
\usepackage{verbatim}
\usepackage{hyperref}

\usepackage{graphicx}
\usepackage{dcolumn,float}
\usepackage{amsmath,amssymb}
\usepackage{subfigure}

\newcommand{\be}{\begin{eqnarray}}

\newcommand{\ee}{\end{eqnarray}}

\newcommand{\qq}{\begin{eqnarray}}
\newcommand{\qqq}{\end{eqnarray}}

\newcommand{\re}[1]{(\ref{#1})}
\newcommand{\eref}[1]{Eq.~(\ref{#1})}

\newcommand{\nm}{\,\mathrm{nm}}

\newcommand{\beg}{\begin{equation}}
\newcommand{\en}{\end{equation}}

\newcommand{\eps}{\epsilon}
\newcommand{\lam}{\lambda}

\usepackage{dcolumn,float}
\usepackage{amsmath,amssymb}
\begin{document}
\title{BCS superconductivity in  metallic nanograins: Finite-size corrections, low energy excitations, and
robustness of shell effects}
\author{Antonio M. Garc\'{\i}a-Garc\'{\i}a}
\affiliation{CFIF, IST, Universidade T\'{e}cnica de Lisboa, Av. Rovisco Pais, 1049-001 Lisboa, Portugal}
\author{Juan Diego Urbina}
\affiliation{Institut f\"ur Theoretische Physik, Universit\"at Regensburg, 93040 Regensburg, Germany}
\author{Emil A. Yuzbashyan}
\affiliation{Center for Materials Theory, Rutgers University, Piscataway, New Jersey 08854, USA}
\author{Klaus Richter}
\affiliation{Institut f\"ur Theoretische Physik, Universit\"at Regensburg, 93040 Regensburg,Germany}
\author{Boris L. Altshuler}
\affiliation{Physics Department, Columbia University, 538 West 120th Street, New York, NY 10027, USA}
\begin{abstract}
We combine the BCS self-consistency condition, a semiclassical expansion for the spectral density and interaction matrix elements to describe analytically how the superconducting gap depends on the size and shape of a 2d and 3d superconducting grain.
 In chaotic grains mesoscopic fluctuations of the matrix elements lead to a smooth dependence of the order parameter on the excitation energy.  In the integrable case we find shell effects i. e. for certain values of the electron number $N$ a small change in $N$ leads to large changes
in the energy gap.
With regard to possible experimental tests we provide a detailed analysis of the dependence of the gap on the coherence length and the robustness of shell effects under small geometrical deformations.

\end{abstract}


\maketitle

\newcommand{\bb}{\boldsymbol{\beta}}
\newcommand{\ba}{\boldsymbol{\alpha}}

Finite size effects are well documented \cite{baduri} in fermionic interacting systems such as atomic nuclei and atomic clusters.
It is also well established \cite{blo,blo1} that the more symmetric the system is,
the stronger are these corrections.
For instance, the existence of magic numbers
 signaling the presence of a particularly stable nucleus has its origin in the gap between the ground state
and the first excited states caused by the high degree of symmetry of the system.

In the field of mesoscopic superconductivity, the study of finite size effects also has a long history. Already fifty years ago, Anderson  noted \cite{ander2} that superconductivity should break down  in small metallic grains when the single particle level spacing at the Fermi energy
is comparable to the bulk
 superconducting gap. In the sixties the size dependence of the critical temperature and the superconducting gap were studied in  for a rectangular grain in \cite{parmenter} and for a nanoslab in\cite{blat}. Thermodynamical properties of  superconducting grains were investigated in \cite{muhl}. Results of these papers are restricted to rectangular grains, and superconductivity is described by the Bardeen, Cooper, and Schriffer (BCS) theory  \cite{BCS}.

The experiments by Ralph, Black, and Tinkham in the mid nineties \cite{tinkham} on Al nanograins of typical size $L \sim 3-13\, \mathrm{nm}$  showed  that the excitation gap
is sensitive to even-odd effects. More recently it has been observed \cite{qsize} that the critical temperature of superconducting ultra-thin lead films oscillates when the film thickness is slightly increased. These results have further stimulated the interest in ultrasmall superconductors \cite{qsize1,fazio,heiselberg,peeters,fomin,richardson}.
For instance,
pairing, not necessarily BCS, in a harmonic oscillator potential was investigated in \cite{heiselberg}.
The critical temperature and the superconducting gap for a nanowire were reported in \cite{peeters} by solving numerically the  Bogoliubov - de Gennes equations. In \cite{fomin}
the superconducting gap and low energy excitation energies in a rectangular grain were computed numerically within the Richardson model \cite{richardson}.
Shell effects in superconducting grains with radial symmetry were studied theoretically in \cite{shell,kresin,usprl}.
Recent experimental results \cite{nmat} in semispherical Sn nanograins have confirmed that shell effects induce strong deviations in the energy gap with respect to the bulk limit. Strong fluctuations of the energy gap as a function of the system size have been observed with a maximum enhancement of about $60 \%$ for sizes $\sim 10$nm.   
Mesoscopic corrections to the BCS energy gap were also considered in \cite{leboeuf,shuck}.

We note that if the mean single particle level spacing is larger than the bulk superconducting gap,
the BCS formalism breaks down. However, an analytical treatment is still possible \cite{duke}
with the help of an exactly solvable model introduced by Richardson \cite{richardson} in the context of nuclear physics.
In particular, finite-size corrections to the predictions of the BCS theory have been recently studied in [\citealp{ML}--\citealp{yuzbashyan}].

Despite this progress, a theory that accounts for all relevant mesoscopic effects in superconducting grains has not emerged so far. The Richardson model
alone cannot provide the foundation for such a theory as it does not allow for mesoscopic spatial fluctuations of the single particle states.
In the present paper, for the particular cases of chaotic and rectangular shaped grains, we develop such a theory based on the BCS theory and semiclassical
techniques. This formalism permits a systematic analytical evaluation of the low energy spectral properties of superconducting nanograins in terms of their size and shape.
Leading finite size corrections to the BCS mean field can also be taken into account in our approach, see
\cite{usprl} for further details. Results for 3d grains were also previously published in \cite{usprl}. Here we discuss both the 2d and 3d cases as well as provide a more detailed account of the
techniques utilized. Moreover, we study the dependence of the mesoscopic BCS order parameter (superconducting gap) on the coherence length, and the
robustness of shell effects.

For chaotic grains, we show that the  order parameter is a universal function of the single particle energy, i.e. it is independent of the particular details of the grain. The mesoscopic fluctuations of the matrix elements of the two-body interactions between single particle eigenstates  are responsible for most of the deviations from the bulk limit.
For integrable grains,  we find that the superconducting gap is strongly sensitive to shell effects. Namely, a
small modification of the grain size or number of electrons inside  can substantially affect its value. 
Throughout the paper we study clean (ballistic) grains. The mean field potential is thus an infinite well of the
form of the grain. We restrict ourselves to system sizes such that the mean level spacing around the Fermi energy is smaller than the bulk gap, so that the
BCS formalism is still a good approximation.
For the superconducting Al grains studied by Tinkham and coworkers \cite{tinkham}, this
corresponds to sizes $L >5\, \mathrm{nm}$.

Our results are therefore valid in the region, $k_F L \gg 1$ (limit of validity of the semiclassical approximation \cite{ander2,ML}), $\delta/\Delta_0 < 1$ (limit of validity of the BCS theory), and  $l\gg\xi \gg L$ (condition of quantum coherence). Here $k_{F}$, $\xi=\hbar v_F/\Delta_0$, $l$, $\delta$,  $\Delta_0$  are the Fermi wave vector,  the superconducting coherence length, the coherence length of the single particle problem, the average single particle level spacing, and the bulk gap. The Fermi velocity  is  $v_F=\hbar k_{F}/m$.
Conditions $k_F L \gg 1$ and $\delta/\Delta_0 < 1$  hold for Al grains  of size $L\gtrsim 5\nm$. Further, in Al grains $\xi \approx 1600  \nm$ and $l > 10^4  \nm$  at temperatures $T \leq 4K$ \cite{albreak}. Therefore, the above region is well accessible to experiments.

\section{The superconducting gap in the BCS theory}
Throughout the paper pairing between electrons is described by the BCS Hamiltonian,
$$
H=\sum_{n\sigma}\eps_n c_{n\sigma}^\dagger c_{n\sigma}-\sum_{n,n'}I_{n,n'} c_{n\uparrow}^\dagger c_{n\downarrow}^\dagger c_{n'\downarrow}c_{n'\uparrow},
$$
where $c_{n\sigma}$ annihilates an electron of spin $\sigma$  in state $n$,
\be
\label{Inn}
I_{n,n'} \equiv I(\eps_{n},\eps_{n'})=\lam V \delta \int \psi^2_n(\vec{r})\psi_{n'}^{2}(\vec{r})d\vec{r}
\ee
are matrix elements of a short-range electron-electron interaction, $\lam$ is the BCS coupling constant.  $\psi_{n}$ and $\eps_n$ are the eigenstates and eigenvalues   of a free particle of effective mass $m$   in a clean grain of volume (area) $V$ ($A$). Eigenvalues $\eps_n$ are measured from the actual Fermi energy $\eps_F$ of the system. In this notation the mean level spacing is $\delta = 1/\nu_{\rm  TF}(0)$, where $\nu_{{  \rm TF}}(0)$ is the spectral density at the Fermi energy
 in the Thomas-Fermi approximation.

The BCS order parameter is defined as
$$
\Delta_n\equiv \Delta(\eps_n)= \sum_n  I_{n,n'} \langle c_{n'\uparrow}^\dagger c_{n'\downarrow}^\dagger \rangle.
$$
Within BCS theory, it is determined by the following self-consistency equation\cite{MaLee}:
\be
\label{gap1}
\Delta_{n} = \frac{1}{2}\sum_{|\eps_{n'}|<\eps_D}
\frac{\Delta_{n'}I_{n,n'}}{\sqrt{\eps_{n'}^2+\Delta_{n'}^2}},
\ee
where $\eps_D$ is the Debye energy. This result is obtained in the grand canonical approximation \cite{BCS}.   Note that, the BCS order parameter $\Delta_n$ is an explicit function of the single-particle energy $\eps_{n}$ since the matrix elements $I(\eps,\eps')$ are energy dependent.

Introducing the exact density of single-particle states $\nu(\eps')=\sum_{n'} \delta(\eps'-\eps_{n'})$, one can
write Eq. (\ref{gap1}) in integral form,
\begin{equation}
\label{gap}
\Delta(\eps)=\frac{1}{2}\int_{-\eps_D}^{\eps_D}\frac{ \Delta(\eps')I(\eps,\eps') }{\sqrt{ {\eps'}^2+\Delta^2(\eps')} }\nu(\eps')d\eps'.
\end{equation}
The gap equation (\ref{gap}) will be the main subject of our interest. As soon as the order parameter
$\Delta(\eps)$ is known, the low lying (single-particle) excitation spectrum, $E=\sqrt{\Delta(\eps)^2+\eps^2}$,
is also determined.

In the large volume (area) limit, the spectral density, to leading order,  is  given by the Thomas-Fermi expression
\begin{equation}
\label{conG}
\nu_{{\rm TF}}(\eps')=2 \times
\begin{cases}
\frac{{V}}{4 \pi^{2}}\left(\frac{2 m}{\hbar^{2}}\right)^{3/2}\sqrt{\eps'+\eps_{F}}, & {\rm \ for \ 3d} \\
\frac{{A}}{4 \pi}\left(\frac{2 m}{\hbar^{2}}\right), & {\rm \ for  \ 2d},
\end{cases}
\end{equation}
where the factor two in front stands for spin degeneracy. In addition, in the bulk limit the matrix  elements (\ref{Inn}) for chaotic grains are simply
$I(\eps,\eps')= \lam \delta$ as a consequence of quantum ergodicity.
The gap is then energy independent $\Delta(\eps) = \Delta_{0}$, and \eref{gap1} yields the BCS bulk result,
\begin{equation}
\label{eq:deltab}
\Delta_{0} = 2\eps_D{\rm e}^{-\frac{1}{\lambda}}.
\end{equation}
As the volume of the grain decreases,   both $\nu(\eps')$ and $I(\eps,\eps')$ deviate from the bulk limit. In this region a more general approach to solve Eq. (\ref{gap}) is needed.

Since we are interested in the regime of many particles ($\nu_{{  \rm TF}}(0)\eps_{F}\gg 1$), an appropriate tool is the semiclassical approximation in general and periodic orbit theory \cite{gut} in particular (see the Appendix for an introduction).
These techniques yield closed expressions for $\nu(\eps')$ and $I(\eps,\eps')$
in terms of quantities from the classical dynamics of the system, which allows us to calculate analytically the resulting superconducting gap. Such explicit expressions for the superconducting gap enable us to study deviations from the BCS theory, the spatial dependence of the gap, and the relevance of shell effects in realistic, not perfectly symmetric grains. 

Our general strategy can be summarized as follows:
\begin{enumerate}
\item Use semiclassical techniques to compute the spectral density  $\nu(\eps')=\sum_{n'} \delta(\eps'-\eps_{n'})$ and
$I(\eps,\eps')$  as series in the small parameter $1/ k_F L$, where $k_{F}$ is the Fermi wave-vector and $L \simeq V^{1/3} (\simeq A^{1/2})$ is the linear size of the grain (section \ref{Semapp} and Appendix).

\item Solve the BCS gap equation (\ref{gap1}) order by order in $1/k_FL$ (Section \ref{Solgap}).


\item Study the impact of small deformations of the shape of a symmetric grain on the gap   in realistic models of the grain (Section \ref{Idvsre}).
\end{enumerate}

Finally we stress that all the parameters in our model $\lambda, k_F, \eps_D, \eps_F$ are the actual parameters that characterize the material at a given grain size and not necessarily the ones at the bulk limit.  

\section{Semiclassical approximation for the density of states and interaction matrix elements.}
\label{Semapp}
The first step to solve the gap equation is to find explicit expressions for the spectral density  $\nu(\eps')$ and the interaction matrix elements
$I(\eps,\eps')$  as series in a small parameter $1/ k_F L$. While the semiclassical approximation for the spectral density has been known for a long time \cite{gut}, the calculation for the matrix elements has only recently attracted some attention \cite{mele,usprl}. Here we state the results and refer the reader to the Appendix for details.
\subsection{Spectral density}
In the semiclassical approximation (see Appendix~\ref{appA}), the spectral density is given by
\begin{equation}
\label{gesum}
\nu(\eps')\simeq \nu_{{ \rm TF}}(0)\left[1+\bar{g}(0)+\tilde{g}_{l}(\eps')\right],
\end{equation}
with a  monotonous  ${\bar g}(\eps')$ and oscillatory  ${\tilde g}(\eps')$ (as functions of system size) parts. The notation $\bar {g} (\eps = 0)$ means that $\bar{g} $ is evaluated at the Fermi energy. This contribution is given by the Weyl expansion \cite{baduri},
\begin{equation}
\label{WE}
\bar{g}(0)=
\begin{cases}
\pm \frac{{S}\pi}{4k_{F}{V}}+\frac{2{\cal C}}{k_{F}^{2}{V}}, & {\rm 3d}, \\
\pm \frac{{\cal L}}{2k_{F}{A}}, & {\rm 2d},
\end{cases}
\end{equation}
for Dirichlet ($-$) or Neumann ($+$) boundary conditions. In Eq. (\ref{WE}), ${S}$ is the surface area of the 3d cavity and ${\cal C}$ its mean curvature, while ${\cal L}$ is the perimeter in the 2d case.

The oscillatory contribution to the density of states is given by the Gutzwiller trace formula \cite{gut},
\begin{equation}
\label{Gcao}
\tilde{g}_l(\eps')= \Re
\begin{cases}
\frac{2\pi}{k_{F}^{2}{ V}}\sum_{p}^{l}A_{p}{\rm e}^{i\left[k_{F}L_{p}+\beta_{p}\right]}{\rm e}^{i\frac{\eps'}{2\eps_F}k_{F}L_{p}} & {\rm 3d}, \\
\frac{2}{k_{F}{A}}\sum_{p}^{l}A_{p}{\rm e}^{i\left[k_{F}L_{p}+\beta_{p}\right]}{\rm e}^{i\frac{\eps'}{2\eps_F}k_{F}L_{p}} & {\rm 2d}.
\end{cases}
\end{equation}
The summation over classical periodic orbits ($p$) with length $L_{p}$ only includes orbits shorter than the quantum coherence length $l$ of the single-particle problem. The semiclassical amplitude $A_p$ and phase $\beta_{p}$ in Eq. (\ref{Gcao}) can also be computed explicitly using the knowledge of periodic orbits. As was mentioned previously the parameters $k_F$ and $\eps_F$ in the above expressions refer to the Fermi wavevector and Fermi energy of the system at a given grain size. Within the free Fermi gas approximation it is possible to relate the bulk Fermi energy with the one at a given finite size by simply inverting the relation \be
\frac{1}{2}N=\int^{\mu}\nu(\eps)d\eps \ee where $\nu(\eps)$ is the spectral density and $N$ is the number of particles.    
\subsection{Matrix elements}
The calculation of the interaction matrix elements $I(\epsilon,\epsilon')$ is more complicated as it requires information about classical dynamics beyond periodic orbits. For a chaotic cavity the final result (Appendix~\ref{appB}),
\begin{eqnarray}
\label{Isum1}
&&I(\eps,\eps')= \\
&&
\begin{cases}
\frac{\lambda}{{V}}\left[1+\bar{I}^{{\rm short}}_{3d}(0)-\frac{\pi^{2}{S}^{2}}{16k_{F}^{2}{V}^{2}}+\bar{I}^{{\rm long}}_{{\rm dg}}(0,\eps - \eps')\right]& {\rm 3d}, \\
& \\
\frac{\lambda}{{A}}\left[1+\bar{I}^{{\rm short}}_{2d}(0,\eps - \eps')+\bar{I}^{{\rm long}}_{{\rm dg}}(0,\eps - \eps')\right]& {\rm 2d},
\end{cases}
\nonumber
\end{eqnarray}
has two types of contributions. Identical pairs of short classical trajectories hitting the boundary once give
\begin{equation}
\label{Isum2}
\begin{cases}
\bar{I}^{{\rm short}}_{3d}(0)=\frac{\pi\cal{S}}{4k_{F}{V}} & {\rm 3d}, \\
 & \\
\bar{I}^{{ \rm short}}_{2d}(0,\eps-\eps')=\frac{{\cal L}}{k_{F}{A}}\left[C'+\frac{{\rm Si}(4k_{F}L)}{\pi}\right] & {\rm 2d}, \\
 +\frac{{\cal L}}{2\pi k_{F}{A}}\left[{\rm Ci}\left(\frac{4(\eps-\eps')k_{F}L}{\eps_F}\right)-{\rm Ci}(\frac{2(\eps-\eps')}{\eps_F})\right]
\end{cases}
\end{equation}
with $C'=0.339...$ a numerical constant given in the Appendix, and ${\rm Ci}(x)$ the cosine-integral function.

In the so-called diagonal approximation (see Appendix~\ref{appB}) the contribution of longer classical trajectories is
\begin{equation}
\label{Isum3}
\bar{I}^{{\rm long}}_{{\rm dg}}(\eps_F,\eps - \eps')=
\begin{cases}
\frac{1}{{V}}\Pi_{l}\left(\frac{\eps - \eps'}{\eps_F}\right) & {\rm 3d}, \\
\frac{1}{{A}}\Pi_{l}\left(\frac{\eps - \eps'}{\eps_F}\right) & {\rm 2d},
\end{cases}
\end{equation}
where
\begin{equation}
\label{pi}
\Pi_{l}(w)=\int\sum_{\gamma}^{l}D_{\gamma}^{2}\cos{\left[wk_{F}L_{\gamma}(\vec{r})\right]}d\vec{r}
\end{equation}
is an integrated sum over trajectories $\gamma(\vec{r})$ starting and ending at position $\vec{r}$. As detailed in Appendix~\ref{appB}, due to the ergodicity of the chaotic classical systems, in the limit $l \gg L$, Eq. (\ref{pi}) simplifies to
\begin{equation}
\label{Isum4}
\Pi_{l \gg L}(w)=
\begin{cases}
\frac{4 \pi^{2}}{k_{F}^{3}}\frac{\sin{\left(wk_{F}l\right)}}{w} & {\rm 3d}, \\
\frac{4}{k_{F}^{2}}\frac{\sin{\left(wk_{F}l\right)}}{w} & {\rm 2d}.
\end{cases}
\end{equation}

For integrable grains there is no universal expression for $I(\eps,\eps')$. We restrict ourselves to the rectangular geometry where to a good approximation the matrix elements are energy independent.

Using the knowledge of $\nu(\eps')$ and $I(\eps,\eps')$ as series  in $1/k_{F}L$, we solve the gap equation (\ref{gap}) in different situations of interest. The resulting gap function, in general, depends the single-particle energy $\eps$, the size of the system, and the number of particles (or, equivalently, Fermi energy $\eps_F$).

\section{Solution of the gap equation in the semiclassical regime}
\label{Solgap}

In this section we solve the gap equation Eq. (\ref{gap}) for $\Delta(\eps)$.
For a rectangular box in two and three dimensions the gap equation is algebraic, since $\Delta(\eps)=\Delta$ is energy independent.  In the chaotic case, however, we get an integral equation due to the energy dependence of the interaction matrix elements. As we will see, both cases can be solved analytically order by order in $1/k_FL$.
\begin{figure}
\includegraphics[width=0.9\columnwidth,clip,angle=0]{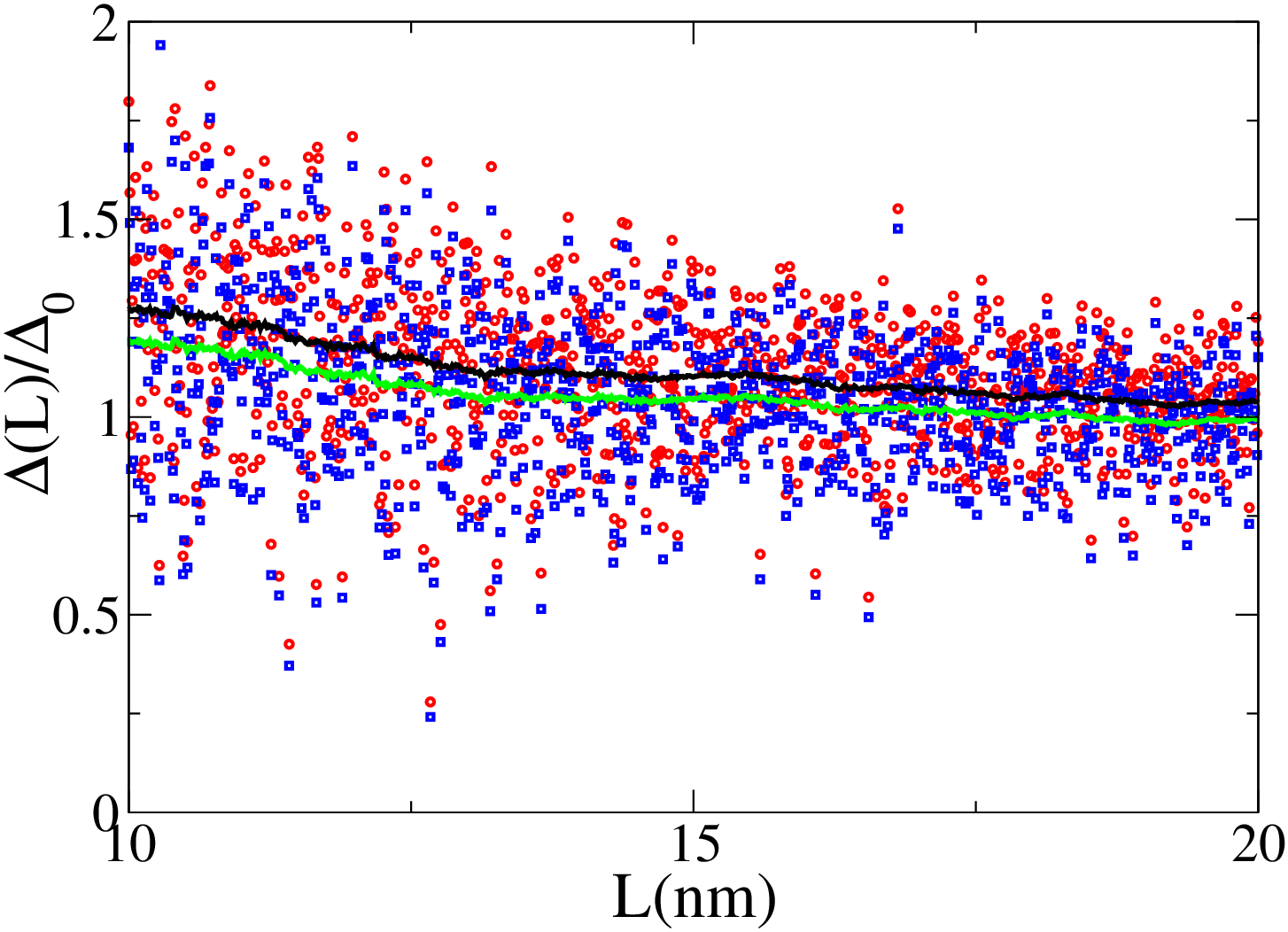}
\includegraphics[width=0.9\columnwidth,clip,angle=0]{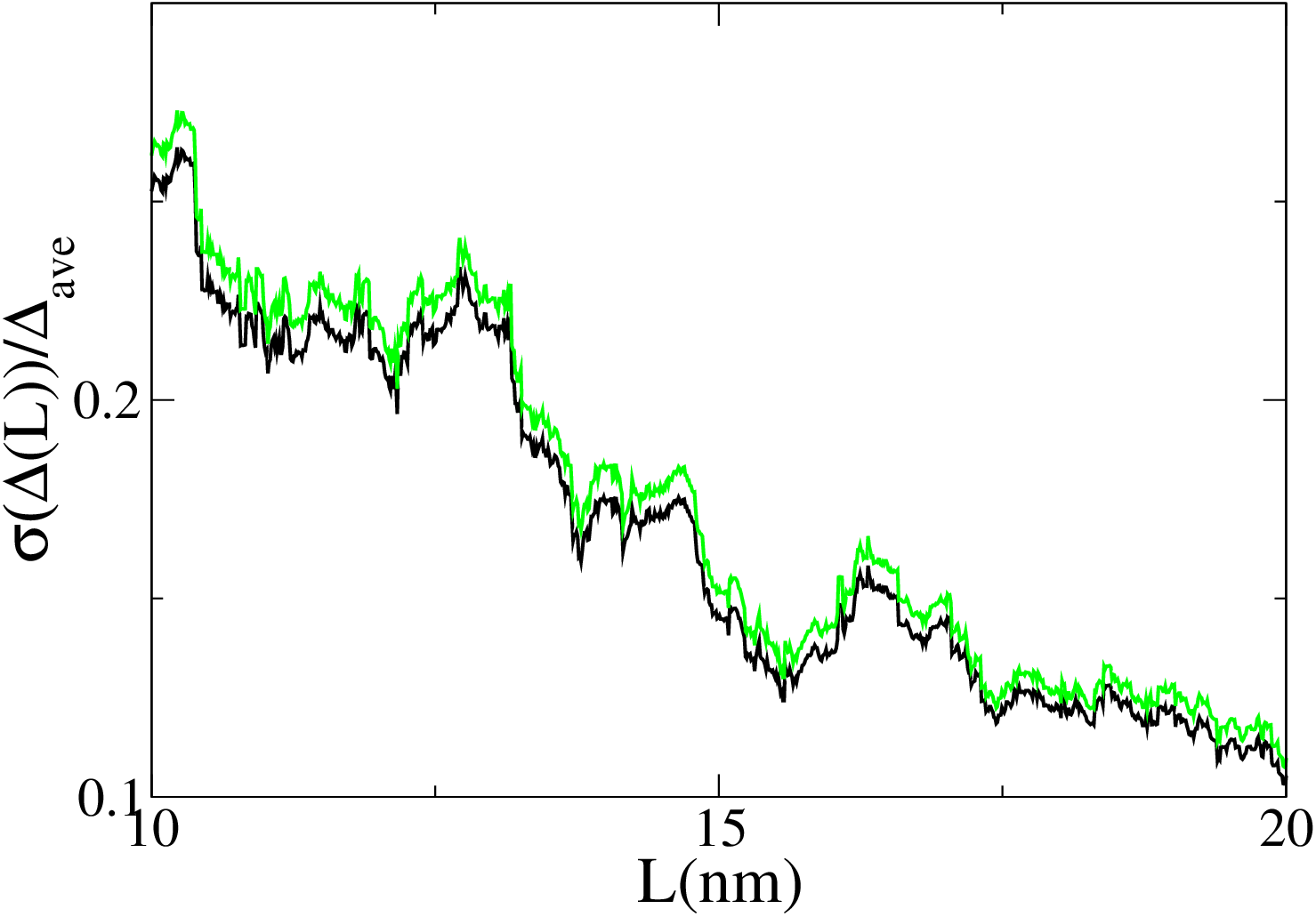}
\vspace{3mm}
\caption{Upper figure:
The energy gap $\Delta$ in units of the bulk gap $\Delta_{0}$ for a cubic grain of side $L$ with $\lambda=0.3$, $\eps_D = 32$meV, $\eps_F \approx 11.65$eV, $k_F = 17.5{\rm nm}^{-1}$ as a function of the grain size $L$. The chemical potential was computed exactly as a function of $N$ by inverting
 the relation $
\frac{1}{2}N=\int^{\mu}\nu(\eps)d\eps$ where $\nu(\eps)$ is the spectral density. Similar results (not shown) are obtained for other values of $\lambda$.  
Red circles stand for the exact numerical solution of the gap equation Eq.(\ref{gap1}) with matrix elements Eq.(\ref{Ire}). The black curve is its average value $\Delta_{ave}$. Blue squares is the numerical solution of Eq.(\ref{gap1}) for trivial matrix elements $I(\eps,\eps') =1/V$. The green curve is its average value. Lower figure: we represent the standard deviation of the gap $\sigma(L)$ in units of the average gap $\Delta_{ave}$, a typical estimation of the average fluctuation, as a function of the grain size. The black (green) curve is the typical deviation for the case of non trivial matrix elements given by Eq.(\ref{Ire}) ($I(\eps,\eps') =1/V$).   
 As can be observed, in the region $\delta/\Delta_0 \gg 1$ (Al $L \gg 6nm$), in which our semiclassical formalism is applicable, the non-trivial matrix element Eq.(\ref{Ire}) does not modify substantially the average gap or the typical fluctuation. We note that the average fluctuation (see also figure 2) is in reasonable agreement with the theoretical prediction, $\frac{\tilde \Delta }{\Delta_0} \approx \sqrt{\frac{\pi\delta}{4\Delta_0}}$ \cite{leboeuf}.}
\label{gapN}
\end{figure}
\subsection{Rectangular box in two and three dimensions}
For the rectangular box the matrix elements are \be 
\label{Ire}
I(\eps,\eps') = \prod_{i=x,y,z}(1+\delta_{\eps_i,\eps_i'}/2)/V \ee where $\eps_i \propto k_i^2$, $ p_i = \hbar k_i$ is the conserved momentum in the $i = x,y,z$ direction and here $\delta$ stands for Kronecker's function. We first investigate the role of these matrix elements on the energy gap.
Qualitatively we expect an enhancement as $I(\eps,\eps') > 1/V$. This enhancement should not be large for $\delta/\Delta_0 \ll 1$ as the spectrum of a rectangular grain has only accidental degeneracy, namely, $\eps_i =\eps_i'$ typically implies that $i=i'$. For a perfectly cubic grain the enhancement is expected to be larger due to level degeneracy although they will still relatively small for $\delta/\Delta_0 \ll 1$. The numerical results of Fig. \ref{gapN} (upper) for the gap as a function of the grain size confirm this prediction. We compare the cases of trivial matrix elements $I(\eps,\eps') \approx 1/V$ and Eq.(\ref{Ire}) (see caption for details).  In the region in which our results are applicable $\delta \ll \Delta_0$ ($L \gg 6$nm) the enhancement of both the gap average (upper plot) and fluctuations (lower plot) due to Eq.(\ref{Ire}) is small. We note that in the numerical calculation the chemical potential is not the bulk Fermi energy but it is computed exactly for each grain size (see caption). This induces an additional enhancement of the average gap with respect to the bulk limit $\Delta_0$.

Since we are mainly interested in the study of gap fluctuations (see below) we neglect in the rest of this section the non trivial part of Eq.(\ref{Ire})  ($I(\eps,\eps') \approx 1/V$). Therefore to a good approximation the gap does not depend on energy, $\Delta(\eps)=\Delta$, and satisfies the equation,
\begin{equation}
\label{Gb3}
\frac{2}{\lambda}=\int_{-\eps_D}^{\eps_D}\frac{1+\bar{g}(0)+\tilde{g}_{l}(\eps')}{\sqrt{\eps'^{2}+\Delta^{2}}}d\eps',
\end{equation}
where $\bar{g}(0)$ for a 3d rectangular box is given by Eq. (\ref{WE}) without the curvature term.

Using Eq.~(\ref{WE}) for $\bar{g}(0)$ and Eq. (\ref{Gcao}) for $\tilde{g}_{l}(\eps')$ (from now on we drop the subscript $l$ to simplify the notation), and taking into account the scaling of each contribution with $1/k_FL$ as described in the Appendix, we look for a solution of the gap equation (\ref{Gb3}) for the 3d case in the following form:
\begin{equation}
\label{Db3}
\Delta=\Delta_{0}(1+f^{(1)}+f^{(3/2)}+f^{(2)}),
\end{equation}
where $f^{(n)}\propto 1/(k_FL)^{n}$. Substituting $\Delta$ into Eq. (\ref{Gb3}), expanding in powers of $1/k_FL$, and equating the coefficients at each power,  we obtain an explicit expression for $f^{(i)}$
\begin{eqnarray}
\label{db23}
\lambda f^{(1)}&=&\left[\bar{g}(0)+\frac{\lambda}{2}\int_{-\eps_D}^{\eps_D}\frac{\tilde{g}^{(3)}(\eps')}{\sqrt{\eps'^{2}+\Delta_{0}^{2}}}d\eps'\right], \nonumber \\
\lambda f^{(3/2)}&=&\sum_{i,j\neq i}^{3}\frac{\lambda}{2}\int_{-\eps_D}^{\eps_D}\frac{\tilde{g}^{(2)}_{i,j}(\eps')}{\sqrt{\eps'^{2}+\Delta_{0}^{2}}}d\eps',
\end{eqnarray}
\begin{eqnarray}
\label{db4}
\lambda f^{(2)}&=&\sum_{i}^{3}\frac{\lambda}{2}\int_{-\eps_D}^{\eps_D}\frac{\tilde{g}^{(1)}_{i}(\eps')}{\sqrt{\eps'^{2}+\Delta_{0}^{2}}}d\eps' \nonumber \\
&+&f^{(1)}\left(f^{(1)} -\bar{g}(0)\right) \\
&-&f^{(1)}\sum_{i}\frac{\Delta_{0}^{2}}{2}\int_{-\eps_D}^{\eps_D}\frac{\tilde{g}^{(1)}_{i}(\eps')}{(\eps'^{2}+\Delta_{0}^{2})^{3/2}}d\eps',\nonumber
\end{eqnarray}
where ${\tilde g^{(k)}} \propto \left(k_F L \right)^{-k}$ denotes the oscillating part of the spectral density. Explicit expressions for $\tilde{g}^{(k)}$,  $\tilde{g}^{(k)}_i$, and $\tilde{g}^{(k)}_{i,j}$ for a rectangular box in terms of periodic orbits  can be found  in the Appendix and also in Ref.~\cite{baduri}.

Equations (\ref{db23}) and (\ref{db4}) can be further simplified by the following argument. After we express  $\tilde{g}^{(3)},\tilde{g}^{(2)}$ and $\tilde{g}^{(1)}$ in terms of a sum over periodic orbits, the integration over $\eps'$ can be explicitly performed. The resulting expression is again an expansion in terms of periodic orbits with two peculiarities: a) the spectral density is evaluated at the Fermi energy and b)
in the limit $\eps_D \gg \Delta_0$ the contribution of an orbit of period $L_p$ is weighted with the function
\be
\label{Wsum}
W(L_p/\xi) = \frac{\lambda}{2} \int_{- \infty}^{\infty} \frac{\cos(L_p t/\xi)}{\sqrt{1+t^2}}dt.
\ee
This cutoff function is characteristic of the BCS theory as opposed to the smoothing due to temperature or inelastic scattering (recall that in this paper we assume that the
single-particle coherence length $l$ is much larger than superconducting coherence length $\xi$). In a similar fashion, the last term in $f^{(2)}$ is weighted with
\begin{eqnarray}
W_{3/2}(L_p/\xi)& =&\frac{\Delta_{0}^{2}}{2} \int_{- \infty}^{\infty}\frac{\cos(L_p t/\xi)}{(1+t^2)^{3/2}} dt.\nonumber
\end{eqnarray}
The effect of $W_{3/2}(L_p/\xi)$ is, again, to  exponentially suppress the contribution of periodic orbits longer than $\xi$. Therefore the sum over periodic orbits in the definition of the spectral density
is effectively restricted to orbits with lengths of the order or smaller than the
superconducting coherence length $\xi$.

Following standard semiclassical approximations, we introduce $\tilde{g}_{\xi}(0)$ as a spectral density evaluated at the Fermi energy  with a cutoff function that suppresses the contribution of orbits of length $L_{p} > \xi$.
With these definitions, we get
\begin{eqnarray}
\label{db234}
\lambda f^{(1)}&=&\left[\bar{g}(0)+\tilde{g}^{(3)}_{\xi}(0)\right], \nonumber \\
\lambda f^{(3/2)}&=&\sum_{i,j\neq i}^{3}\tilde{g}^{(2)}_{i,j \xi}(0), \\
\lambda f^{(2)}&=&\sum_{i}^{3}\tilde{g}^{(1)}_{i \xi}(0) \nonumber \\
&+&f^{(1)}\left[f^{(1)} -\bar{g}(0)-\sum_{i}^{3}\tilde{g}^{(1)}_{i \xi}(0)\right].\nonumber
\end{eqnarray}
Eq. (\ref{db234}) is our final result for the finite size corrections to the gap function for a 3d rectangular box. As expected, it is expressed in terms of classical quantities such as the volume, surface, and periodic orbits of the grain. 

\begin{figure}
\includegraphics[width=0.9\columnwidth,clip,angle=0]{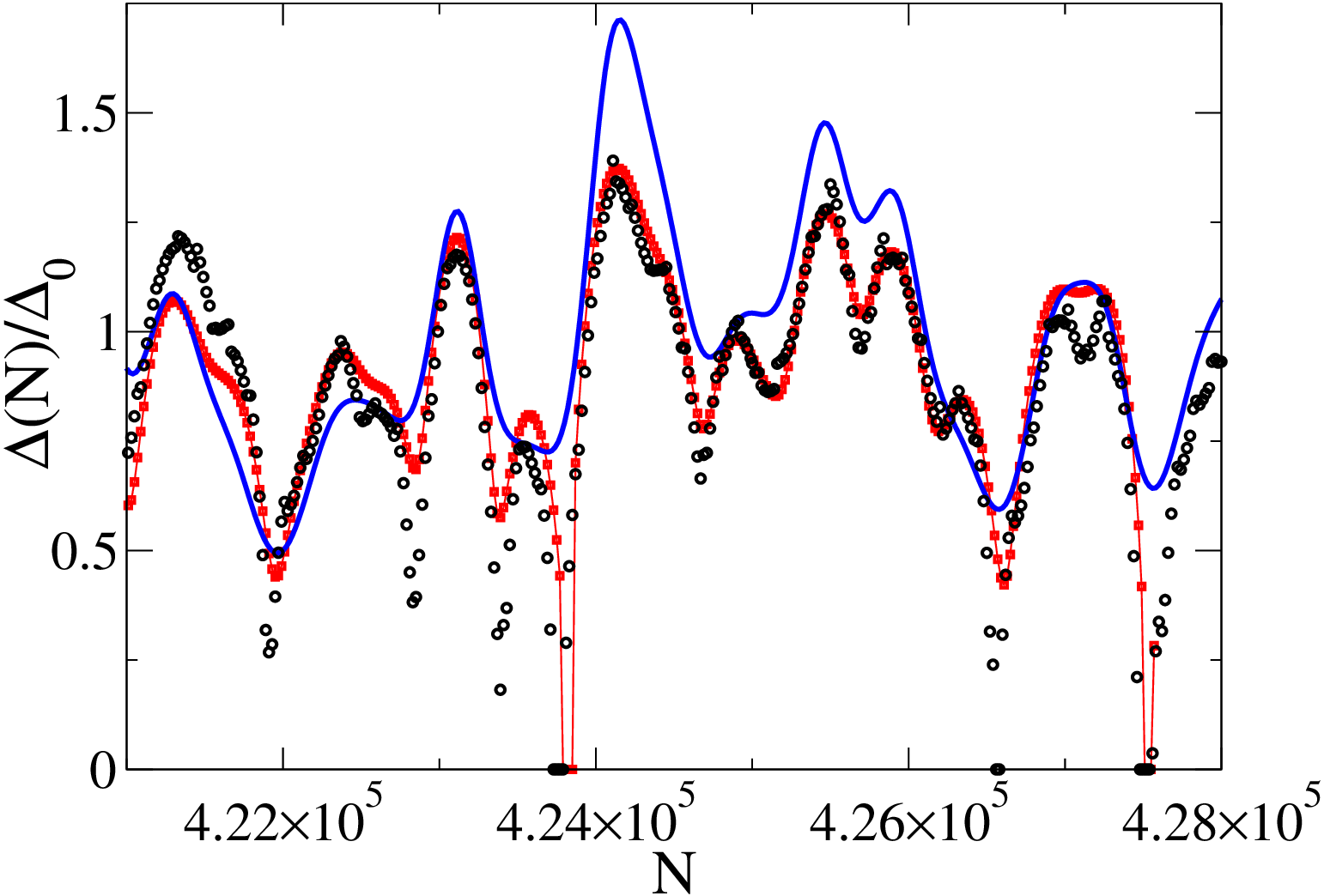}
\vspace{3mm}
\caption{The energy gap $\Delta$ in units of the bulk gap $\Delta_{0}$ for a cubic grain with $\lambda=0.3$, $\eps_D = 32$meV, $\eps_F \approx 11.65$eV, $k_F = 17.5{\rm nm}^{-1}$ as a function of the number of particles $N$ ($L \approx 13.23-13.32$nm) inside of the grain. The solid line is the analytical prediction from (\ref{Db3}) and (\ref{db234}). Black circles ( red squares) are results from a numerical evaluation of the gap equation using the exact (semiclassical Eq. (\ref{gesum})) spectral density. The semiclassical formalism provides an excellent description of the exact numerical results for the gap. We stress that, for the sake of simplicity, it has been assumed that $I = 1/V$.   }
\label{gapN1}
\end{figure}

In Fig. \ref{gapN1} we compare the analytical expression for the gap (\ref{Db3}) and (\ref{db234}) (solid blue line) to the numerical solution of the gap equation using the exact one-body spectrum (circles) and the semiclassical prediction for the spectral density (red squares). It is observed that the analytical expression for the gap is in fair agreement with the exact numerical results. Moreover it is also clear from the figure  that the semiclassical formalism provides an excellent description of the numerical results. We note that the small differences observed for small values of the gap are a consequence of the finite $l \sim 50R$ single particle coherence length entering in the semiclassical expression of the spectral density Eq.(\ref{gesum}). Since our motivation here is test the validity of the semiclassical formalism we are assuming for simplicity that the chemical potential is fixed at the bulk Fermi energy.   

The following argument can shed light on our results.
The density of states cannot  be pulled out of the energy integration in the gap equation (\ref{Gb3}) unless it is smoothed. However, this is exactly what our result Eq. (\ref{db234}) means, since truncating the sums is equivalent to smoothing the energy dependence. We conclude that our result Eq. (\ref{Gb3}) should be similar to the  standard BCS solution in the bulk, $\Delta_{0}=2\eps_D{\rm e}^{-1/\lambda}$, with the substitution $\lambda \to \lambda(1+\bar{g}(0)+\tilde{g}_{\xi}(0))$. Indeed, an expansion of this expression in $1/k_FL$ gives exactly Eq. (\ref{db234}).

In order to simplify notation from now on we will drop the subscript $\xi$  in the spectral density $\tilde{g}_{\xi}$ smoothed by the cutoff function $W(L_{p}/\xi)$. In  2d we find,
\begin{equation}
\label{Db3b}
\Delta=\Delta_{0}(1+f^{(1/2)}+f^{(1)}),
\end{equation}
with
\begin{eqnarray}
\lambda f^{(1/2)}&=&\tilde{g}^{(2)}_{1,2}(0) \nonumber \\
\lambda f^{(1)}&=&\bar{g}(0)+\sum_{i=1,2}\tilde{g}^{(1)}_{i}(0)\\ &+&\frac{1-\lambda}{\lambda}\left[\tilde{g}^{(2)}_{1,2}(0)\right]^{2}. \nonumber
\end{eqnarray}
The sums implicit in $\tilde{g}_{i}, \tilde{g}_{i,j}$ are smoothly truncated by the same weight function $W(L_p/\xi)$.  Similar to the 3d case, the above result can also be obtained by expanding the bulk expression for the gap  with the full density of states in $1/(k_F L)$. We note that, contrary to the 3d case, in 2d grains, oscillatory contributions to the density of states are of leading order.

\subsection{3d chaotic cavity}

The energy dependence of the interaction matrix elements, $I(\eps,\eps')$, in this case is given by Eqs.~(\ref{Isum1}--\ref{Isum4}), i.e.
\begin{eqnarray}
I(\eps,\eps')& = & \frac{\lambda}{V}\left[1+ \frac{\pi S}{4k_F V} - \frac{\pi^2 S^2}{16 k_F^2 V^2} + \frac{1}{V}\Pi_{l}\left(\frac{\eps-\eps'}{\eps_F}\right)\right], \nonumber
\end{eqnarray}
where
\begin{equation}
\label{picao}
\Pi_{l}(w) = \frac{4\pi^2}{k_F^3}\frac{\sin(k_F l \omega)}{\omega}.
\end{equation}
The details of the calculation  based on the semiclassical approximation for  Green's functions can be found in Appendix~\ref{appB}.

The above expression for  $I(\eps,\eps')$ together with the semiclassical expression for the spectral density (\ref{Gcao}) are the starting point for the calculation of the superconducting order parameter.
The energy dependence of the matrix elements implies a gap equation of integral type and, most importantly, that the order parameter itself depends on the energy. Based on the $1/k_FL$ dependence of the different contributions to $I(\eps,\eps')$, we write
\begin{equation}
\label{deltacha3}
\Delta(\eps)=\Delta_{0}\left[1+f^{(1)}+f^{(2)} + f^{(3)}(\eps)\right]
\end{equation}
for a 3d chaotic grain. Substituting this expression into the gap equation \re{gap} and comparing powers of $1/k_FL$, we get  a simple algebraic equation for $f^{(1)}$ with the solution
\begin{equation}
\label{d1cao}
\lambda f^{(1)}=(1\pm 1)\frac{{\cal S} \pi}{4k_{F}{V}}.
\end{equation}
It shows that for Dirichlet (-) boundary conditions, the superconducting order parameter for a chaotic 3d cavity does not have mesoscopic deviations of order $1/k_F L$. This suppression is a hallmark of the chaotic case and appears due to the fluctuations of the interaction matrix elements. It can be also found by substituting $\lambda \to \lambda (1+{\cal S} \pi/4k_{F}{V})$ into \eref{eq:deltab}, which accounts only for the surface contribution to the density of states, and expanding the modified $\Delta_0$ to first order in $1/k_FL$ \cite{shuck}.

The second order correction reads
\begin{equation}
\label{d2cao}
\lam f^{(2)}=  \frac{2{\cal C}}{k_{F}^{2}{V}}+
2\left(\mp 1+ \frac{1\pm 1}{\lambda}\right)\left(\frac{\pi {\cal S}}{4 k_{F}V}\right)^{2}+\tilde{g}(0),
\end{equation}
with
\begin{equation}
\tilde{g}(0) =
\frac{2\pi}{k_{F}^{2}V}\sum_{p}A_{p}W(L_p/\xi)\cos(k_{F}L_{p}+\beta_p),
\label{g0}
\end{equation}
where the contribution of periodic orbits $L_p$ longer than the coherence length $\xi$ is exponentially suppressed.

Equating terms of order $(k_F L)^{-3}$, we obtain for $f^{(3)}(\eps)$ an integral equation of the form $f^{(3)}(\eps)=h(\eps)+\int K(\eps') f^{(3)}(\eps') d\eps'$, which is solved with the ansatz $f^{(3)}(\eps)=h(\eps)+c$, where $c$ is a constant. We obtain
\begin{eqnarray}
\label{fx}
f^{(3)}(\eps)&=&\frac{\pi \lambda
\delta}{\Delta_0}\left[\frac{\Delta_0}{\sqrt{\eps^{2}+\Delta_0^2}}+\frac{\pi}{4}\right].
\end{eqnarray}
Note that a)  since $\delta/\Delta_0 \ll 1$ is an additional small parameter the contribution \re{fx} can be comparable to
lower orders in the expansion in $1/k_FL$ and
 b) the order parameter $\Delta(\eps)$ has a maximum at the Fermi energy ($\eps=0$) and
decreases on an energy scale $\eps\sim\Delta_0$ as one moves away from the Fermi level.  One can also show that mesoscopic corrections given by Eqs.~(\ref{d1cao}), (\ref{d2cao}) and \re{fx} always enhance $\Delta(0)$ as compared to the bulk value $\Delta_0$.
A couple remarks are in order: a) the energy dependence of the gap is universal in the sense that it does not depend on
specific grain details, b) the matrix elements $I(\eps,\eps')$ play a crucial role, e.g. they are responsible for most of the deviation from  the bulk limit. Finally we briefly address the interplay of mesoscopic fluctuations and 
parity effects (see \cite{usprl} for a more detailed account). The Matveev-Larkin (ML) parity parameter $\Delta_p$ \cite{ML}, a experimentally accessible observable, accounts for even-odd asymmetries in ultrasmall superconductors. While the ML parameter coincides with the
 standard superconducting gap in the bulk limit, in \cite{ML} it was found that its leading finite size correction is given by
\be
\label{ope}
\Delta_p \equiv E_{2N+1} - \frac12\big(E_{2N}+E_{2N+2}\big)=\Delta(0) - \frac{\delta}{2},
\ee
where  $E_N$ is the ground state energy for a superconducting grain with $N$ electrons. 

We see that these corrections to the BCS mean-field approximation are comparable to mesoscopic fluctuations but have an opposite sign. For Al it seems that mesoscopic corrections are larger than those coming from (\ref{ope}).

\subsection{2d chaotic cavities}

In this section we study a 2d superconducting chaotic grain of area $A$, perimeter $\cal{L}$, and linear size $L=\sqrt{A}$. Our starting point is the gap equation \re{gap} together with the semiclassical expressions for the spectral density, Eqs.~\re{WE} and \re{Gcao}, and the matrix elements, $I(\eps,\eps')$, Eqs.~(\ref{Isum1}--\ref{Isum4}), namely
\begin{eqnarray}
\label{I2dc}
I(\eps,\eps')&& = \frac{\lambda}{A}\left[1+
\frac{{\cal L}}{k_{F}{A}}\left[C'+\frac{{\rm Si}(4k_{F}L)}{\pi}\right]+\right. \nonumber \\ &&\left.
\frac{{\cal L}}{2\pi k_{F}{A}}\left[{\rm Ci}\left(\frac{4(\eps - \eps')k_{F}L}{\eps_F}\right)-{\rm Ci}\left(\frac{2(\eps - \eps')}{\eps_F}\right)\right] \right. \nonumber \\ &&\left. + \Pi_{l}\left(\frac{\eps-\eps'}{\eps_F}\right) \right],
\end{eqnarray}
where $C' \approx 0.339 \ldots$ and ${\rm Si}(x), {\rm Ci}(x)$ are the sine and cosine integral functions, respectively. For $l \gg L$, the chaotic classical dynamics leads to a universal form for the function $\Pi_{l}(w)$,
\be
\Pi_{l}(w) = \frac{4}{k_F^2}\frac{\sin(k_F l \omega)}{\omega}.
\ee
As in the 3d case, the energy dependence of matrix elements implies that the equations to be solved for the gap are of integral type, and that the gap itself is energy dependent. However, unlike the 3d case, we have logarithmic corrections coming from the contribution of the matrix elements. Based on the expansion in powers of $1/k_FL$ of the spectral density and $I(\eps,\eps')$ [see also Eqs.~\re{Iz} and (\ref{Ifin})] we propose for a 2d chaotic grain the expansion
\begin{equation}
\label{Delta2DD1}
\Delta(\eps)=\Delta_0\left[1+f^{(\log)}+f^{(1)} + \pi^{-1}f^{(2)}(\eps)\right].
\end{equation}
Following the same steps to solve the gap equation as in the 3d case, we get to leading order,
\be
\lambda f^{(\log )}=\frac{{\cal L}\log{2k_{F}L}}{2\pi k_{F}{A}}.
\ee
Similar logarithmic corrections to residual interactions in 2d chaotic quantum dots in the Coulomb Blockade regime were reported in Ref.~\onlinecite{mele}.

The next order correction is given by
\be
\lambda f^{(1)}= (C' \pm 1)\frac{{\cal L}}{2k_{F}{\cal A}} + \tilde{g}(0),
\ee
with $(-)$ for Dirichlet and $(+)$ for Neumann boundary conditions, respectively. The truncated spectral density  $\tilde{g}(0)$ is defined as in the 3d case, with semiclassical amplitudes corresponding to 2d systems.

Finally, the energy dependent correction to the gap in 2d chaotic grains, $f^{(2)}(\eps) $ is given by the same function (\ref{fx}) as in 3d grains.

We note that a) in 2d the leading finite size contribution comes from the interaction matrix elements, not from the spectral density, b)  finite size effects are stronger than in  3d and the leading correction does not vanish for any boundary condition, c) since effectively there are two expansion parameters $1/k_FL \ll 1$ -- assuring the validity of the semiclassical approximation-- and $\delta/\Delta_0 < 1$ -- in order to apply the BCS formalism-- it can happen that in a certain range of parameters the contribution $f^{(2)}(\eps)$ is dominant.

\begin{figure}[ht]
\includegraphics[width=0.95\columnwidth,clip,angle=0]{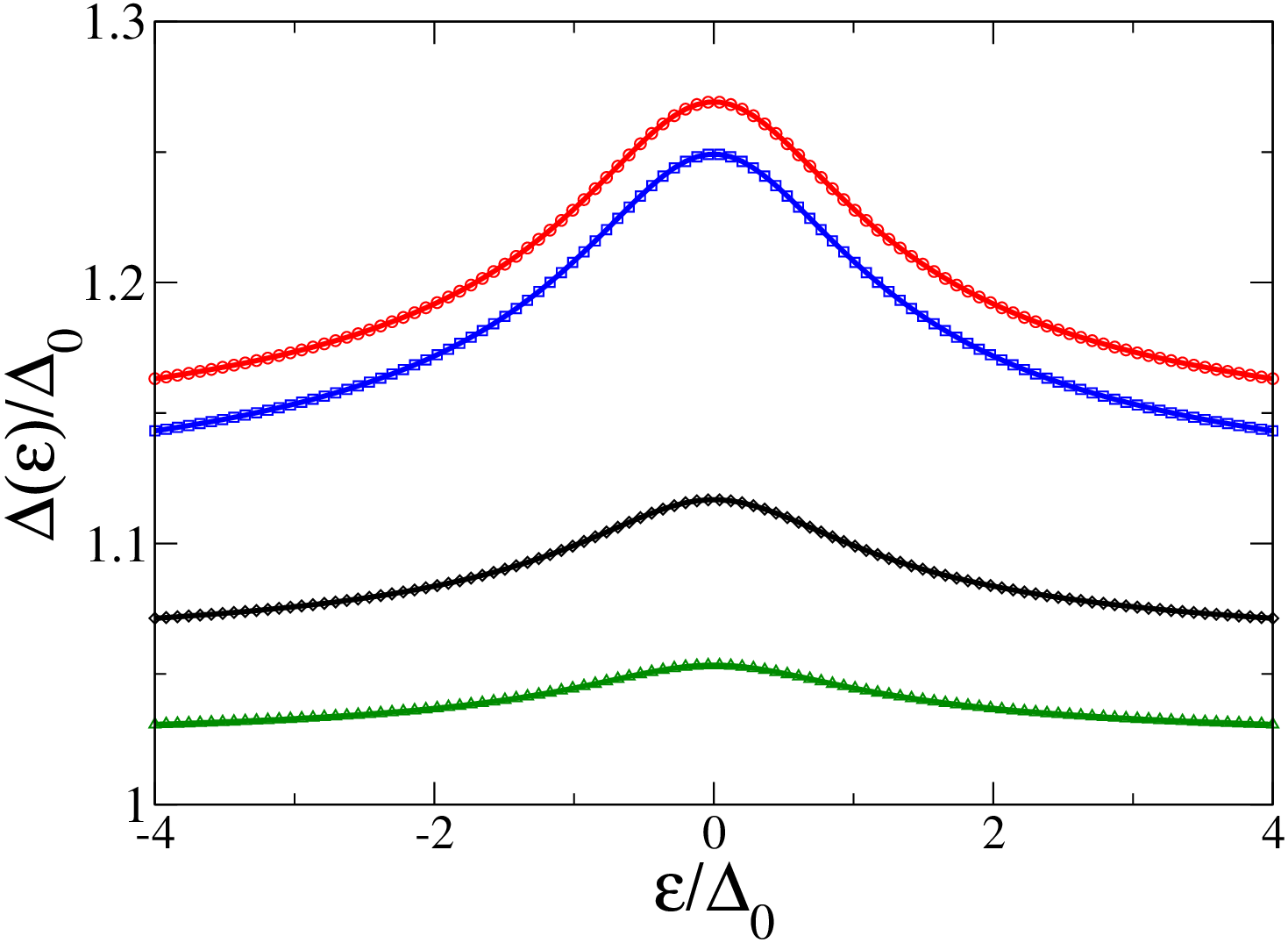}
\vspace{3mm}
\caption{Superconducting order parameter $\Delta(\eps)$, Eq. (\ref{Delta2DD1}), in units of the bulk gap $\Delta_0$ for 2d chaotic Al grains
($k_F = 17.5 \nm^{-1}, \delta = 7279/N, \Delta_0 \approx 0.24 meV$) as a function of the energy $\eps$ with respect to the Fermi level $\eps = 0$. 
Different curves correspond to grain sizes (top to bottom) and boundary conditions:
$L=6\nm, k_F L = 105, \delta/\Delta_0 = 0.77)$ (Dirichlet and Neumann boundary conditions), $L= 8\nm, k_F L = 140, \delta/\Delta_0 = 0.32$ (Dirichlet), and  $L= 10\nm, k_F L = 175, \delta/\Delta_0 = 0.08$  (Dirichlet).  The leading contribution comes from the energy dependent matrix elements $I(\eps,\eps')$.}
\label{gap2D}
\end{figure}

In Fig. \ref{gap2D} we plot the gap as a function of the energy in units of the bulk gap $\Delta_0$ for Al grains ($k_F \approx 17.5 \mbox{nm}^{-1}$, $\lambda \approx 0.18$, and $\delta \approx 7279/N \mbox{ meV}$ where $N$ is the number of particles), of different sizes
$L$. Note the single peak at the Fermi energy. For the smallest grains the leading contribution is  $f^{(2)}(\eps)$. This is yet another indication that the matrix elements play a dominant role
in the finite size effects in superconducting metallic grains.

\section{Enhancement of superconductivity in nanograins: Ideal versus real grains}
\label{Idvsre}

According to the findings of previous sections the superconducting gap is an oscillating function of the
system size and the number of electrons inside the grain. Even for grains with $N \sim 10^4 - 10^5$ electrons considerable deviations from
the bulk limit are observed.
For a fixed grain size, the deviations from the bulk limit are the larger the more symmetric the grain is. This is a typical shell effect similar to that found in other fermionic systems, such as nuclei and atomic clusters \cite{baduri}.
These shell effects have their origin in the geometrical symmetries of the grain. Symmetries induce degeneracies in the spectrum and, consequently, stronger fluctuations in the spectral density. The superconducting  gap is enhanced if the Fermi energy is in a region of level bunching (large spectral density). Likewise, if the
Fermi energy is close to a shell closure (small spectral density) the superconducting gap will be much smaller than  in the bulk limit.

Therefore, thanks to shell effects, one can adjust the gap value by adding or removing few electrons
 in such a way that the Fermi energy moves into a region of high or
low spectral density.  In fact, shell effects in metallic grains of different geometries have recently attracted considerable attention \cite{fomin,peeters,shell,kresin,moro,friedel}.
A superconducting spherical shell and a
rectangular grain were studied numerically in Ref.~\onlinecite{fomin}. A similar analysis was carried out in
Ref.~\onlinecite{peeters} for a nanowire.
A qualitative analysis of a spherical superconductor was reported in Ref.~\onlinecite{shell}.

Discrepancies with experiments are
expected because factors such as decoherence,  deformations of the shape of the grain, and surface vibrational modes are not taken into account in the theoretical analysis.
In this section we discuss the impact  of small deformations of the grain and of decoherence effects that shorten the coherence length. We will see that weakly deformed grains can be modeled as symmetric ones but with an effective
coherence length that incorporates the details of the deformation.  The semiclassical formalism utilized in this paper is especially suited to tackle this problem.

\subsection{Superconductivity and shell effects}

We study the dependence of the gap on the number of electrons $N$ inside the grain and compare the
gap between two grains with slightly different degree of symmetry.
We focus on 3d rectangular grains where deviations from the bulk results are expected to
be larger. In this case the chemical potential can be computed exactly as a function of $N$ by inverting
 the relation 
 \be
\frac{1}{2}N=\int^{\mu}\nu(\eps)d\eps \ee where $\nu(\eps)$ is the spectral density. 
\begin{figure}[ht]
\includegraphics[width=0.95\columnwidth,clip,angle=0]{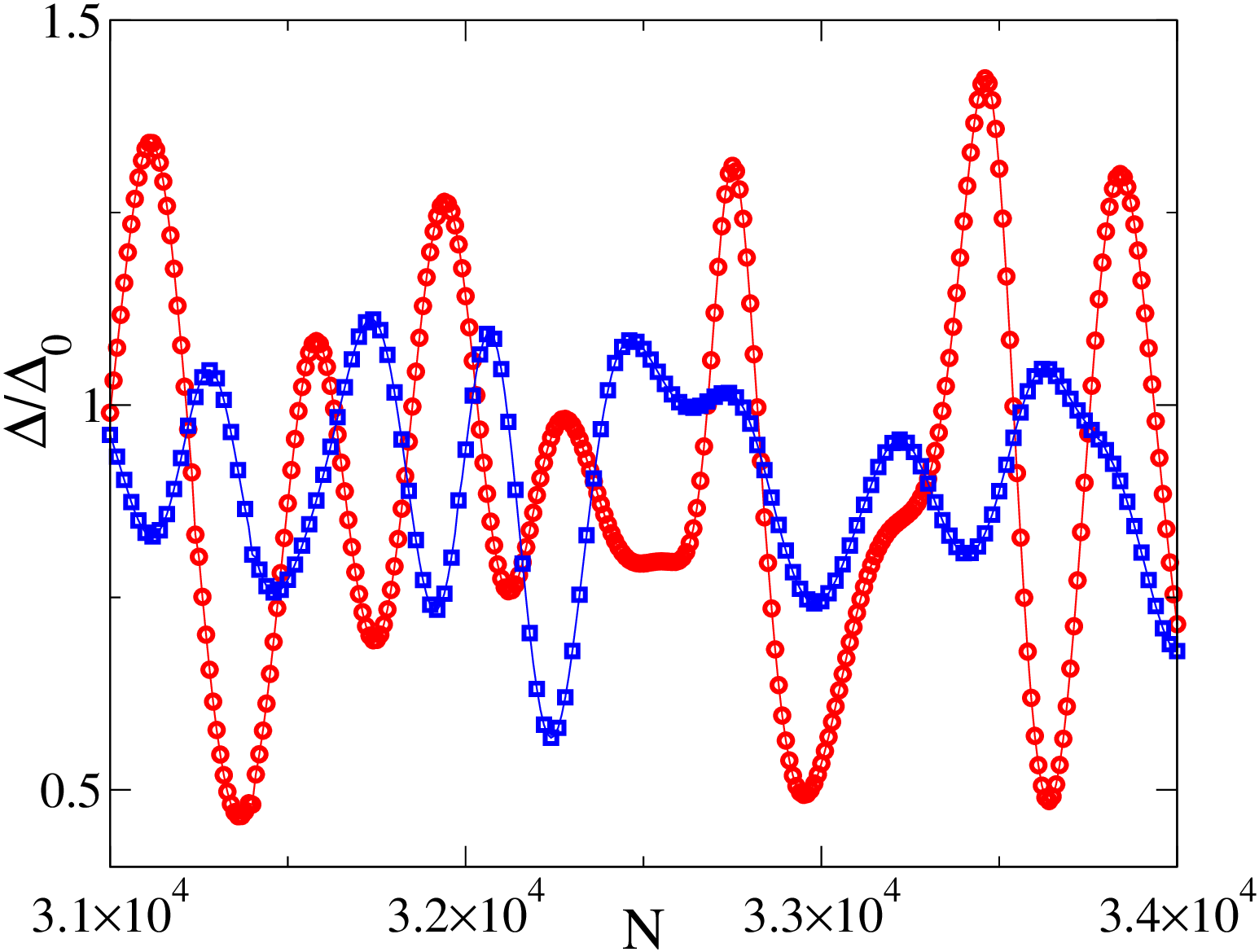}
\label{fig4}
\vspace{3mm}
\caption{
The superconducting gap $\Delta$ in units of $\Delta_{0} \approx 2.286$meV, as a function 
of the particle number $N$ for a cubic (circles), of side $L$, and a parallelepiped-shaped ($1.0288:0.8909:1.0911$) (squares) grain.
 Fluctuations are on average stronger in the cubic grain due to its larger symmetry.
The parameters utilized are $\lambda = 0.3$, $\eps_D =32$meV, $\eps_F \approx 11.85$eV, $k_F = 26 {\rm nm}^{-1}$. The energy gap was obtained by solving Eq.(\ref{Gb3}) with the semiclassical expression of the spectral density given by by Eqs. (\ref{Gint},\ref{Gint3},\ref{Gint2}) and a single particle coherence length $l \sim 12 {\rm L}$. }
\label{Fluct}
\end{figure}

As it is shown in Fig. 1 matrix elements does not affect the gap oscillations.   
Therefore we can solve the gap equation (\ref{gap}) following the
steps of section \ref{Solgap} with the spectral density given by Eqs. (\ref{Gint},\ref{Gint3},\ref{Gint2}) and $I \approx 1/V$.
The spectral density depends on the cutoff, namely, on the number of periodic orbits taken into account. This cutoff is set by the single-particle coherence length $l$. Here we take $l \sim 12 L$ where $L$ is the length of the longest side of the
parallelepiped and study the differences between a cubic and a rectangular grain. The cutoff is chosen to be
much larger than the system size in order to observe fluctuations but considerably smaller than the superconducting coherence length $\xi$ in order to accommodate other effects (see below) that might reduce the typical single-particle coherence length in realistic nanograins.
We study a range of $N$ such that the BCS theory is still applicable but deviations from the bulk limit are still important.

In Fig.~\ref{Fluct} we plot $\Delta$, from Eq.(\ref{gap}),
 as a function of $N$ for a cube an a parallelepiped with aspect ratio $1.028:0.89:1.091$. For both settings we observe strong fluctuations with respect to the
bulk value. The fluctuations are clearly stronger in the cubic case since the grain symmetry is larger.
We also observe that a slight modification of the grain size (or equivalently $N$) can result in substantial changes of the gap.
The observed differences between the cube and the parallelepiped are due to the different symmetry of these grains. In the cube the overall
symmetry factor in the spectral density is $\propto N^{1/2}$. The parallelepiped has only two symmetry axis and therefore the symmetry factor $\sim N^{1/3}$. 

In addition to the fluctuations due to periodic orbits, we also expect smooth
corrections to the bulk limit due to the surface and perimeter term of the spectral density.
These corrections will be clearly observed as the coherence length is shortened and the
contribution of periodic orbits is therefore suppressed.

\begin{figure}[ht]
\includegraphics[width=0.95\columnwidth,clip,angle=0]{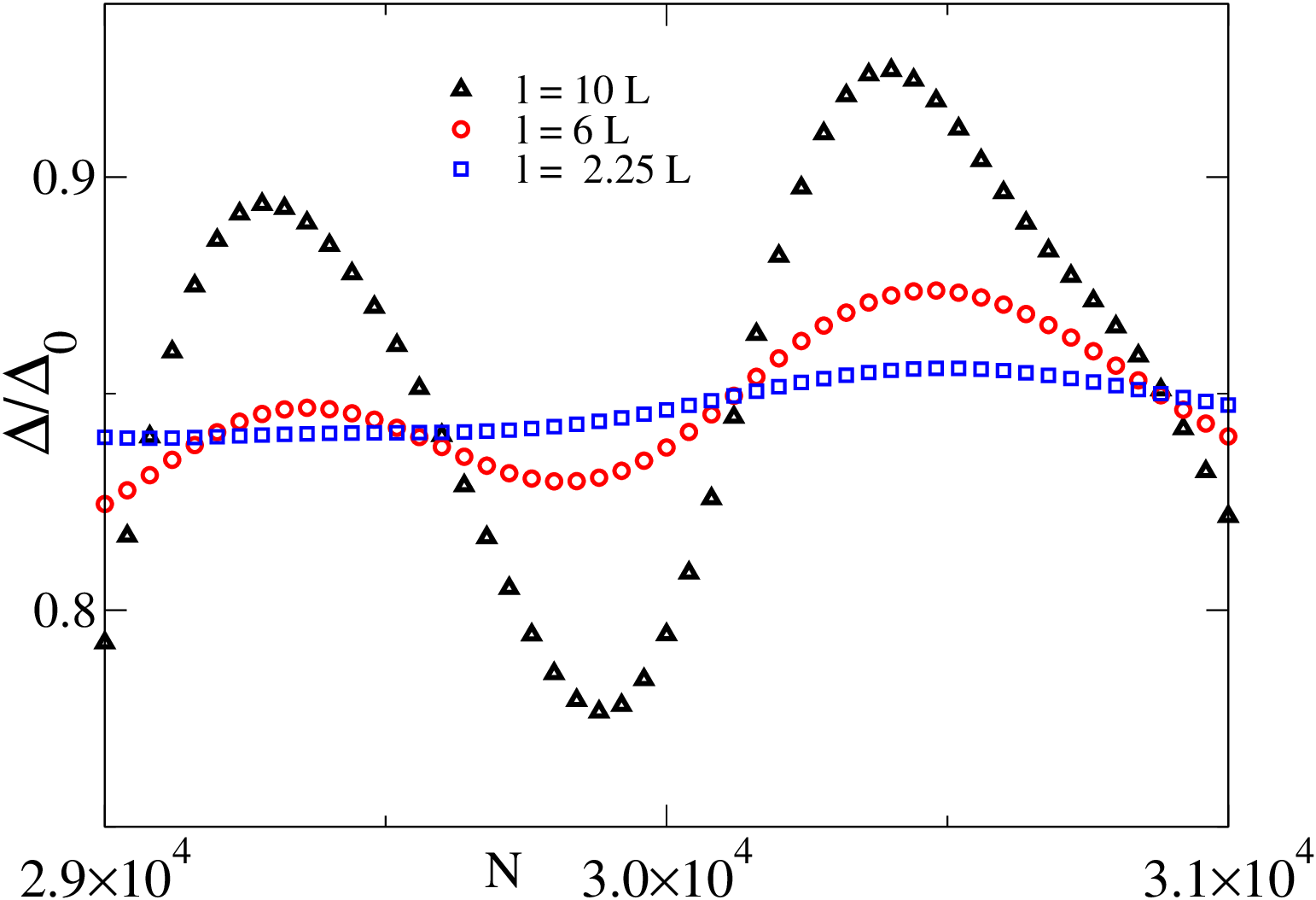}
\label{fig5}
\vspace{3mm}
\caption{Superconducting gap $\Delta$ for a cubic grain (volume $N/181 \nm^3$) for different single particle coherence lengths $l = 2.25L$, $l = 6L$, $l = 10L$ in units of $\Delta_0 \approx 0.228$meV as a function of the number of particles $N$. The parameters utilized are $\lambda = 0.3$, $\eps_D =32$meV, $\eps_F \approx 5.05$eV, $k_F = 18 {\rm nm}^{-1}$. The energy gap was obtained by solving Eq.(\ref{Gb3}) with the semiclassical expression of the spectral density given by Eq.(\ref{gesum}). As the coherence length is reduced
less periodic orbits contribute to the spectral density and fluctuations are smaller. Fluctuations are strongly suppressed for coherence lengths $l \leq 2 L$. In this limit the gap is still smaller than $\Delta_0$ as a consequence of the surface and curvature terms in Eq.(\ref{Gb3}).}
\label{Coher}
\end{figure}

\subsection{Finite size effects in real small grains}

Highly symmetric shapes are hard to produce in the laboratory. It is thus natural to investigate to what extent small deformations from a perfect cubic shape weaken the finite size effects described in previous sections.
For applications it is also important to understand the dependence of the results on the single particle coherence length $l$.
In order to study this dependence, we assume that the superconducting coherence length $\xi$ is the largest length scale in the system. This is the
most interesting region because in the opposite case $l \gg \xi$ the results for the gap (\ref{db234})   are to a great extent independent of $l$. By contrast, in the limit
$\xi \gg l$,
the cutoff (\ref{Wsum}) induced by $\xi$   has little effect as the contribution of periodic
orbits
$L_p \geq \xi$ is already strongly suppressed by the cutoff induced by $l$. If $l \sim \xi$ both cutoffs must be taken into account.

We now address these two related issues. We note that not only the effect of a finite coherence length $l$ but also small deviations from symmetric shapes can be included in our analytical expressions for the gap by adding
an additional cutoff ${\cal D}$ (besides Eq. (\ref{Wsum}))
which suppresses the contribution of periodic orbits longer than ${\cal D}$. The
details of {$\cal D$} depend strongly on the source of decoherence or the type of weak deformation.
 Indeed, in certain cases ${\cal D}$ may modify not only the amplitude but also
the phase of the contribution of the periodic orbit to the trace formula used to compute the
spectral density.
For instance the effect of small multipolar corrections to an otherwise spherical grain \cite{creagh} is modelled by
adding an additional $\cal D$ cutoff in term of a Fresnel integral that smoothly modulates the amplitude and phase of the periodic orbits of the ideal spherical grain.

If the deformation is in the form of small, non overlapping bumps, \cite{Pav98}, the cutoff is exponential and only affects the amplitude.
The numerical value of the cutoff depends on the original grain and is directly related to the typical size of the bump. If the source of decoherence is due to finite temperature  effects, \cite{patricio},
${\cal D} = \frac{L_p/l}{\sinh(L_p/l)}$ with $l$ inversely proportional to the temperature.

In Fig. \ref{Coher} we show the effect of a finite coherence length $l$ in the superconducting cubic grain
investigated previously.
The gap equation Eq. (\ref{gap}) was solved exactly with the
semiclassical spectral density given by Eqs. (\ref{Gint},\ref{Gint1},\ref{Gint2}) and $I=1/V$. For simplicity we use ${\cal D} = \frac{L_p/l}{\sinh(L_p/l)}$ as a cutoff with $l$ now the single particle coherence length.  
This is enough for a qualitative description of the suppression of shell effects as a consequence of decoherence or geometrical deformations.

The cutoff Eq.(\ref{Wsum}), related to the superconducting coherence length, does not affect the calculations as it is much longer ($\sim 1600 \nm$) than the ones employed in Fig. \ref{Coher}. Similar results are obtained if the analytical result (\ref{db234}) is utilized.

As expected, the amplitude is reduced and
 the fine structure of the fluctuations is washed out as the
coherence length is shortened. We
did not observe any gap oscillations with $N$ for $l \geq 2.5 L$. This can be regarded as an effective threshold
for a future experimental verification of shell effects in superconductivity.
Smooth non-oscillatory corrections depending on the ${\cal S}$ (or perimeter ${\cal L}$ in 2d) term in the spectral density are not affected by the coherence length and should be clearly observed in experiments.
Note that $\Delta$ in Fig. \ref{Coher} is, on average,
below $\Delta_0$ even for the maximum $N$ investigated. This is a direct consequence of the negative sign of the surface term in the spectral
density for Dirichlet boundary conditions used in the numerical calculations ($f^{(1)}$ in Eq. (\ref{Db3})).

\section{Conclusions}

We have determined the low energy excitation spectrum, $E=\sqrt{\Delta(\eps)^2+\eps^2}$ of small superconducting grains
 as a function of their size and shape by combining the BCS mean-field approach and semiclassical techniques. For chaotic grains the non-trivial mesoscopic corrections to the interaction matrix elements make them energy dependent, which, in turn,
 leads to a universal smooth energy dependence \re{fx} of the order parameter $\Delta(\eps)$, see Fig~\ref{gap2D}.
 In the integrable (symmetric) case we found that  small changes in the number of electrons can substantially modify
 the superconducting gap, see e.g. Fig~\ref{Fluct}. Due to its potential relevance for experiments, we have investigated how these shell effects decrease (Fig.~\ref{Coher}) when the grain symmetry and/or the single-particle coherence length are reduced.

\begin{acknowledgments}

AMG thanks Jorge Dukelsky for fruitful conversations and acknowledges financial support from FEDER and the Spanish DGI through Project No.
FIS2007-62238. KR and JDU acknowledge useful conversations with Jens Siewert and financial support from the Deutsche Forschungsgemeinschaft (GRK 638). EAY's research was  in part supported by the David and Lucille Packard
Foundation and by the National Science Foundation under Award No. NSF-DMR-0547769.

\end{acknowledgments}

\appendix*
\section{Semiclassical approximation for the density of states and the interaction matrix elements.}

Semiclassical techniques such us periodic orbit theory \cite{baduri} are not a common tool in the study of superconductivity
however they are a key ingredient in our analytical
treatment. In order to solve the gap equation Eq. (\ref{gap}) we first need a closed expression for the spectral density and the interaction matrix elements $I(\eps,\eps')$.
In this Appendix we describe in detail how these quantities are computed using a semiclassical approximation for $1/k_{F}L \ll 1$, where $k_{F}=k(\eps_F) = \frac{\sqrt{2m\eps_F}}{\hbar}$ is the momentum at the Fermi energy $\eps_F$ and $L$ is the  linear system size.  The resulting semiclassical expansion will be organized in powers (possibly fractional) of the small parameter $1/k_F L$.

In order to observe deviations from the bulk limit, the single-particle coherence length  must be larger than the system size, $l \geq L$.
  The time scale, $\tau \approx l/v_F  $, associated with $l$ has a meaning of the lifetime of states near the Fermi energy.  The condition $l \geq L$ means that the Cooper pairs are composed of quasiparticles with a lifetime longer than the flight time through the system.

\subsection{Density of states}
\label{appA}

We start with the analysis of the density of states.
The semiclassical expression for $\nu(\eps)$ for a given grain geometry is already known  in the literature \cite{baduri},
\begin{equation}
\nu(\eps')\simeq \nu_{{ \rm TF}}(0)\left[1+\bar{g}(\eps)+\tilde{g}_{l}(\eps')\right]
\end{equation}
The spectral density gets both monotonous ${\bar g}(\eps)$ and oscillating ${\tilde g}(\eps)$ corrections. The monotonous correction at the Fermi energy is given by the Weyl expansion.
\begin{equation}
\label{GW}
\bar{g}(0)=
\begin{cases}
\pm \frac{{S}\pi}{4k_{F}{V}}+\frac{2{\cal C}}{k_{F}^{2}{V}} & {\rm 3d} \\
\pm \frac{{\cal L}}{2k_{F}{A}} & {\rm 2d}
\end{cases}
\end{equation}
for Dirichlet ($-$) or Neumann ($+$) boundary conditions. In Eq.~(\ref{GW}), $\cal{S}$ is the surface area of the 3d cavity, ${\cal C}$ is its mean curvature, while ${\cal L}$ is the perimeter in the 2d case.

The oscillatory contribution to the density of states is sensitive to the nature of the classical motion. For a system whose classical counterpart is fully chaotic  it is given to the leading order by the Gutzwiller trace formula \cite{gut},
\begin{equation}
\label{gutA}
\tilde{g}_l(\eps')= \Re
\begin{cases}
\frac{2\pi}{k_{F}^{2}{ V}}\sum_{p}^{l}A_{p}{\rm e}^{i\left[k_{F}L_{p}+\beta_{p}\right]}{\rm e}^{i\frac{\eps'}{2\eps_F}k_{F}L_{p}} & {\rm 3d} \\
\frac{2}{k_{F}{A}}\sum_{p}^{l}A_{p}{\rm e}^{i\left[k_{F}L_{p}+\beta_{p}\right]}{\rm e}^{i\frac{\eps'}{2\eps_F}k_{F}L_{p}} & {\rm 2d},
\end{cases}
\end{equation}
where we used $k(\eps')\simeq k_{F}+e'k_{F}/2\eps_F$. The summation is over a set of classical periodic orbits ($p$) of lengths $L_{p}<l$. Only orbits shorter than the quantum coherence length $l$ of the single-particle problem are included. The amplitude $A_{p}$ increases with the degree of symmetry of the cavity \cite{baduri} (see below). In the chaotic case $A_{p}=A_{p}(\eps_F)$ is given by
\begin{equation}
\label{GcaoA}
A_{p}(\eps_F)=\frac{L_{p}}{\left|{\rm det}\left({\bf{M}_{p}}-{\bf{I}}\right)\right|^{1/2}},
\end{equation}
with the monodromy matrix ${\bf{M}_{p}}$ taking into account the linearized classical dynamics around the periodic orbit. The classical flow also determines \cite{baduri}  the topological index $\beta_{p}$ in Eq.~\ref{gutA}.

Note that Eqs~(\ref{gutA}) and (\ref{GcaoA}) indicate that the scaling of $\tilde{g}$ in terms of the small parameter
\begin{equation}
\zeta=1/k_{F}L
\end{equation}
is
\begin{equation}
\label{Gcaoz}
\tilde{g}_l(\eps')\propto
\begin{cases}
\zeta^{2} & {\rm 3d}, \\
\zeta & {\rm 2d}.
\end{cases}
\end{equation}

\subsubsection*{Rectangular grain}

Consider a rectangular box of sides $a_{i}$ with $i=1,\ldots,d$ in $d$ dimensions. For these systems the sum over periodic orbits is exact and given by
 \cite{baduri,blo1},
\begin{equation}
\label{Gint}
\tilde{g}(\eps')=
\begin{cases}
\tilde{g}^{(3)}(\eps')& \\
-\frac{1}{2}\sum_{i}\sum_{j \neq i}\tilde{g}^{(2)}_{i,j}(\eps') & {\rm 3d},\\
+\frac{1}{4}\sum_{i}g^{(1)}_{i}(\eps') & \\
 \\
\tilde{g}^{(2)}_{1,2}(\eps') & \\
-\frac{1}{2}\sum_{i}g^{(1)}_{i}(\eps') & {\rm \ 2d}.
\end{cases}
\end{equation}
Here $\tilde{g}^{(3)}$ is a sum over  families of periodic orbits. Each family is parametrized by three (non simultaneously zero) integers $\vec{n}=(n_{1},n_{2},n_{3})$
\begin{equation}
\label{Gint3}
\tilde{g}^{(3)}(\eps')=\sum_{L_{\vec{n}} \neq 0}^l j_{0}
(k_{F}L_{\vec{n}}+\frac{e'}{2\eps_F}k_{F}L_{\vec{n}})
\end{equation}
where $L_{\vec{n}}=2\sqrt{a_{1}^{2}n_{1}^{2}+a_{2}^{2}n_{2}^{2}+a_{3}^{2}n_{3}^{2}}$ is the length of an orbit in the family and $j_{0}(x)=\sin{x}/x$ is the spherical Bessel function. We see that
\begin{equation}
\label{Gint3z}
\tilde{g}^{(3)} \propto \zeta.
\end{equation}
In the same spirit, $\tilde{g}^{(2)}_{i,j}$ is written as a sum over families of periodic orbits parallel to the plane defined by sides $a_{i}, a_{j}$. In this case the families are labeled by two integers $\vec{n}=(n_{1},n_{2})$ and
\begin{eqnarray}
\label{Gint2}
&& \tilde{g}_{i,j}^{(2)}(\eps')= \\
&&
\begin{cases}
\frac{a_{i}a_{j}\pi}{k_{F}{V}}\sum_{L\vec{n}\neq 0}^l J_{0}\left(k_{F}L_{\vec{n}}^{i,j}+\frac{e'}{2\eps_F}k_{F}L_{\vec{n}}^{i,j}\right) & {\rm 3d} \\
\frac{a_{i}a_{j}}{{A}}\sum_{L_{\vec{n}}\neq 0}^l J_{0}\left(k_{F}L_{\vec{n}}^{i,j}+\frac{e'}{2\eps_F}k_{F}L_{\vec{n}}^{i,j}\right) & {\rm 2d}
\end{cases} \nonumber
\end{eqnarray}
where $L^{i,j}_{\vec{n}}=2\sqrt{a_{i}^{2}n_{1}^{2}+a_{j}^{2}n_{2}^{2}}$ is the length the orbit  $(n_{1},n_{2})$ and $J_0$ is a Bessel function.  Using the asymptotic expression for $J_0$, we find that this contribution scales with $\zeta$ as,
\begin{equation}
\label{Gint2z}
\tilde{g}_{i,j}^{(2)}(\eps')\propto
\begin{cases}
\zeta^{3/2} & {\rm 3d} \\
\zeta^{1/2}& {\rm 2d}.
\end{cases}
\end{equation}
Finally,  for $\tilde{g}^{(1)}_{i}$ we have periodic orbits labeled by a single integer $n$
\begin{equation}
\label{Gint1}
\tilde{g}_{i}^{(1)}(\eps')=
\begin{cases}
\frac{4\pi a_{i}}{k_{F}^{2}{V}}\sum_{L_{n}^{i}}^l\cos{\left(k_{F}L_{n}^{i}+\frac{e'}{2\eps_F}k_{F}L_{n}^{i}\right)} & {\rm 3d} \\
\frac{4a_{i}}{k_{F}{A}}\sum_{n}\cos{\left(k_{F}L_{n}^{i}+\frac{e'}{2\eps_F}k_{F}L_{n}^{i}\right)} & {\rm 2d}
\end{cases}
\end{equation}
with lengths $L_{n}^{i}=2na_{i}$. The dependence on $\zeta$ in this case is
\begin{equation}
\label{Gint1z}
\tilde{g}_{i}^{(1)}(\eps')\propto
\begin{cases}
\zeta^{2} & {\rm 3d} \\
\zeta & {\rm 2d}.
\end{cases}
\end{equation}
It is important to note that depending on the classical dynamics and the spatial dimensionality there are different types of scaling with $\zeta$. The amplitude of the spectral fluctuations increases with the degree of symmetry of the cavity.  It is maximal in spherical
 cavities and minimal in cavities with no symmetry axis \cite{baduri}. The latter typically includes chaotic cavities, namely,
cavities such that the motion of the classical counterpart is chaotic.

This relation between symmetry and fluctuations can be understood as follows.
In grains with
one or several  symmetry axis there exist periodic orbits of the same length. As a result of taking all these degenerate orbits into
account, the amplitude of the spectral density is enhanced by a factor ${\zeta}^{-1/2}$
 for each symmetry axis \cite{bal,blo2}. For instance,  a spherical cavity has three symmetry axis so the
symmetry factor is proportional to ${\zeta}^{-3/2} \gg 1$.
Periodic orbits in chaotic cavities are not in general degenerate
and the symmetry factor is therefore equal to one.
For the range of sizes  $L \sim 5-10 {\rm nm}$ studied in this paper the difference between a chaotic and
an integrable grain can be orders of magnitude.

\subsection{Interaction matrix elements}
\label{appB}

\subsubsection{Semiclassical approximation to the average density}

Unlike the case of the density of states,
there is no general semiclassical  theory for quantities, such as the interaction matrix element $I(\eps,\eps')$, involving the spatial integration of more than two eigenfunctions in clean systems. For integrable systems the ergodic condition,
\be
I(\eps,\eps') = \frac{\lambda}{\Omega}
\label{II}
\ee
with $\Omega =V$ or $A$ in 3d and 2d respectively, is typically not met due to the existence of constants of motion.
The constraints imposed by conservation laws effectively localize the eigenfunctions in a smaller region of the available phase space.

On the other hand,  for chaotic systems Eq.~(\ref{II}) is well justified as a result of the quantum ergodicity theorem \cite{erg1}.
   The vast majority of the eigenfunctions spread almost uniformly over the whole volume (area)  due to the lack of constants of motion besides the energy.
 If the position $\vec{r}$ is far enough from the boundaries, we have
\begin{equation}
|\psi^{2}_{n}(\vec{r})|^{2}=\frac{1}{\Omega}(1+ O(\zeta))
\end{equation}
for almost all states close to the Fermi energy.  In order to evaluate explicitly deviations from Eq.~(\ref{II}), we propose the replacement
\begin{equation}
\label{erg}
|\psi^{2}_{n}(\vec{r})|^{2} \to \langle |\psi(\vec{r})|^{2}\rangle_{\eps_{n}}.
\end{equation}
The average is over a small window of states around $\eps_n$. The width of this window is controlled by an energy scale $\hbar/\tau$ related to the single-particle coherence length $l \approx v_F \tau$. This averaging procedure is justified since eigenfunctions of classically chaotic systems have well defined statistical properties \cite{av1}.


The above average is exactly given by
\begin{eqnarray}
\label{Green}
\langle |\psi(\vec{r})|^{2} \rangle_{\eps}&=& \frac{1}{g(\eps)}\sum_{\eps_{n}}w(\eps-\eps_{n}) |\psi_{e_{n}}(\vec{r})|^{2} \\
&=&\frac{1}{\pi g(\eps)}\int w(\eps')\Im{G(\vec{r},\vec{r},\eps'-\eps+i0^{+})}d\eps' \nonumber
\end{eqnarray}
where $G(\vec{r},\vec{r}',z)$ is the Green function of the non-interacting system at complex energy $z$, $w(x)$ is a normalized window function of width $\hbar / \tau$ centered around $x=0$, and $g(\epsilon)$ is the density of states smoothed by $w(x)$.

Next, we express the Green function as, 
\begin{equation}
\label{GreenI}
G=G^{0}+\tilde{G}.
\end{equation}
where $G^0$ is given by the free propagator
\begin{equation}
\label{Greenf}
G^{0}(\vec{r},\vec{r}',\eps +i0^{+})=
\begin{cases}
-\frac{m}{2 \pi \hbar^{2}} \frac{{\rm e}^{ik(\eps)|\vec{r}-\vec{r}'|}}{|\vec{r}-\vec{r}'|} & {\rm 3d} \\
-\frac{im}{2 \hbar^{2}} H_{0}^{+}(k(\eps)|\vec{r}-\vec{r}'|) & {\rm 2d},
\end{cases}
\end{equation}
and $H_{0}^{+}$ is the Hankel function. The corresponding contribution to the average intensity, obtained by taking the limit $\vec{r} \to \vec{r}'$ of the imaginary part of $G^{0}$ on \ref{Greenf}, is then spatially uniform and given by
\begin{equation}
\label{Int0}
\langle |\psi(\vec{r})|^{2} \rangle_{\eps}^{0}=\frac{1}{\Omega}.
\end{equation}
The effect of such so-called zero-length paths joining $\vec{r}$ with $\vec{r}$ in zero time is then to produce a constant background independent of the position (see, for example \cite{bogo}). This result should not come as a surprise, as zero-length paths are responsible for the leading order terms in the Weyl expansion of the density of states.

In the semiclassical approach \cite{gut} the other part of the Green function, $\tilde{G}$, is expressed in terms of non-zero paths $\gamma$ going from $\vec{r}$ to $\vec{r}$ in a finite time $\tau_{\gamma}$ as,
\begin{equation}
\label{Greenosc}
\tilde{G}(\vec{r},\vec{r},\eps)=\sum_{\gamma}D_{\gamma}{\rm e \ }^{i\left(k_{F}L_{\gamma}+\frac{\eps}{2\eps_F}k_{F}L_{\gamma}+\beta_{\gamma}\right)}.
\end{equation}
 This contribution is responsible of the typical spatial oscillations of the average intensity. The classical properties of each trajectory are encoded in its topological phase $\beta_{\gamma}$ (equal to $\pi/4$ times the number of conjugate points reached by the trajectory) and the smooth function $D_{\gamma}=D_{\gamma}(\vec{r},\vec{r}',\eps_F)|_{\vec{r}=\vec{r}'}$ \cite{baduri,gut},
\begin{equation}
\label{Ds}
D_{\gamma}(\vec{r},\vec{r}',\eps_F)=
\begin{cases}
\frac{1}{k_{F}}\left|{\rm det \ }\frac{\partial^{2}L_{\gamma}(\vec{r},\vec{r}')}{\partial q_{i} \partial q'_{j}}\right|^{1/2} & {\rm 3d}, \\
\sqrt{\frac{2}{ \pi k_{F}}}\left|\frac{\partial^{2}L_{\gamma}(\vec{r},\vec{r}')}{\partial q \partial q'}\right|^{1/2} & {\rm 2d}.
\end{cases}
\end{equation}
Here $q_{i}$ and $q'_{j}$ are local coordinates transverse to the trajectory $\gamma$ at points $\vec{r}$, respectively, and $\vec{r}'$, and $L_{\gamma}(\vec{r},{\vec{r}}')$ is its length. In 3d we have two perpendicular components, while in 2d there is only one. 

After substitution of Eq. (\ref{Greenosc}) into Eq. (\ref{Green}), the integration over energies can be carried out explicitly provided that, in consistency with the stationary phase approximation used to derive the semiclassical Green function, all smooth functions of the energy are evaluated at $\epsilon_{\rm F}$. The resulting Fourier transform of the window function acts as a cut-off for the sum. 
We finally obtain, after using  the expression for the density of states and factorizing the Thomas-Fermi density, 
\begin{equation}
\label{Int}
\langle |\psi(r)|^{2} \rangle_{\eps}=\frac{1}{\Omega}\frac{1+\tilde{R}(\vec{r},\eps)}{1+\bar{g}(\eps_F)+\tilde{g}(\eps)}
\end{equation}
In both 3d and 2d, $\tilde{R}(\vec{r},\eps)$ is simply obtained form the Green function as a sum over classical paths $\gamma(\vec{r})=\gamma$ starting and ending at point $\vec{r}$ with finite lengths $L_{\gamma}(\vec{r})=L_{\gamma}<l$ and actions $S_{\gamma}(\vec{r})=\hbar k(\eps)L_{\gamma}$ \cite{KJD2}
\begin{equation}
\label{Rsem}
\tilde{R}(\vec{r},\eps)=\sum_{\gamma}^{l}D_{\gamma}\cos{\left(k_{F}L_{\gamma}+\frac{\eps}{2\eps_F}k_{F}L_{\gamma}+\beta_{\gamma}\right)}.
\end{equation}

Inspection of Eq.~(\ref{Ds}) shows that $\tilde{R}$ scales as
\begin{equation}
\label{Dz}
D_{\gamma} \propto
\begin{cases}
\zeta& {\rm \ \ 3d} \\
\zeta^{1/2}& {\rm \ \ 2d}
\end{cases}.
\end{equation}
Furthermore, the normalization condition implies
\begin{equation}
\label{trace}
\frac{1}{\Omega}\int \tilde{R}(\vec{r},\eps)d\vec{r}=\bar{g}(\eps)+\tilde{g}(\eps).
\end{equation}
Eq.~(\ref{trace}) can also be used as the definition of the density of states without the Thomas-Fermi contribution.

The separation between smooth, $\bar{g}(\eps) \simeq \bar{g}(0)$, and oscillatory terms $\tilde{g}(\eps)$ in Eq. (\ref{trace}) is as follows. Smooth contributions come from trajectories starting and ending at $\vec{r}$ after hitting the boundary only once $L_{\gamma}< L$. On the other hand, trajectories hitting the boundary more than once will have in general $L_{p}> L$, and their contribution to the spatial integral can be evaluated using the stationary phase approximation to give $\tilde{g}(\eps)$.

Using Eqs. (\ref{GW},\ref{Gcaoz},\ref{Dz}) and (\ref{trace}) the interaction matrix elements have the following semiclassical expansion,
\begin{equation}
\label{Iexp}
I(\eps,\eps')=
\begin{cases}
\frac{\lambda}{{V}}[1+\bar{I}(\eps_F,\eps,\eps') -\frac{{\cal{S}}^{2}\pi^{2}}{16k_{F}^{2}{V}^{2}}] & {\rm 3d},\\
& \\
\frac{\lambda}{{A}}[1+\bar{I}(\eps_F,\eps,\eps')-\frac{{\cal L}}{2k_{F}{A}}+\tilde{g}_l(\eps_F)] & {\rm 2d},
\end{cases}
\end{equation}
where
\begin{equation}
\label{Ismoo}
\bar{I}(\eps,\eps')=\frac{1}{\Omega}\int\tilde{R}(\vec{r},\eps)\tilde{R}(\vec{r},\eps')d\vec{r}.
\end{equation}

\subsubsection{Evaluation of $\bar{I}(\eps,\eps')$}

 As we will see,  $\bar{I}(\eps,\eps')$ is a smooth function of both $\eps$ and $\eps'$; it does not oscillate as rapidly as ${\rm e}^{iS/ \hbar}$ where $S$ is the classical action.
The key point in carrying out the spatial integration in Eq. (\ref{Ismoo}) is the separation of $\tilde{R}=\tilde{R}^{{\rm short}}+\tilde{R}^{{\rm long}}$ into short and long classical trajectories. A similar separation leads to the smooth and oscillatory contributions to the density of states discussed in the previous section. In other words, our approach to evaluating $\bar{I}(\eps,\eps')$ is similar to the Weyl expansion for the density of states.

To calculate $\tilde{R}^{{\rm short}}$, we note that in the regime $l \geq L$, the short trajectories of length $L_{\gamma} < L$ are insensitive to the smoothing, and hence their contribution to the imaginary part of the Green function in (\ref{Green}) can be pulled out from the energy integration. This means that $\tilde{R}^{{\rm short}}$ is simply proportional to the imaginary part of the Green function associated with the short paths.

Following Balian and Bloch \cite{bal} the basic idea of the subsequent calculation is that the boundary of the grain can be locally approximated as a plane in 3d and a straight line in 2d provided that the observation point is close enough to it. Within this approximation, the exact Green function representing a single reflection off an infinite wall can be calculated using the method of images. For the 3d case Eq. (\ref{Ds}) is quantum mechanically exact and gives the same result.

Following this idea, we construct the Green function for an infinite straight boundary by means of the method of images to obtain 
\begin{equation}
\label{Greenshort}
G^{\rm short}(\vec{r},\vec{r}',\eps +i0^{+})=\pm G^{0}(\vec{r},T(\vec{r}'),\eps +i0^{+})
\end{equation}
where the action of the linear operator $T$ is to map the position $\vec{r}'$ into its image point on the other side of the boundary. The plus and minus sign give Neumann and Dirichlet boundary conditions respectively. 

It is easy to see that when $\vec{r} \to \vec{r}'$, the function $G^{\rm short}$ depends only on the distance between $\vec{r}$ and the boundary, which we denote by $x$. Since this distance is still of the order of the system linear size, it is possible to perform the energy average in Eq. (\ref{Green}) with the Green function given by Eq. (\ref{Greenshort}). As a result, 
\begin{equation}
\label{Rshort}
\tilde{R}^{{\rm short}}(\vec{r},\eps)=\pm
\begin{cases}
\frac{\sin{2k(\eps)x}}{2k(\eps)x} & {\rm 3d}, \\
J_{0}(2k(\eps)x) & {\rm 2d},
\end{cases}
\end{equation}

After  $\tilde{R}^{{\rm short}}$ is inserted into Eq. (\ref{Ismoo}), the integral along directions parallel to the plane simply yields a factor of $\cal{S}$ in 3d and $\cal{L}$ in 2d. The integration in the perpendicular direction is naturally truncated at the system linear size $L$. In 3d, using $\int_{0}^{L}=\int_{0}^{\infty}-\int_{L}^{\infty}$, we obtain
\begin{equation}
\label{I3D}
\bar{I}^{{\rm short}}_{3d}(\eps,\eps')=-\frac{\cal{S}}{8k_{F}^{2}LV}+ \\\frac{\pi \cal{S}}{4V}\frac{{\rm Min \ }\left[k(\eps),k(\eps')\right]}{k(\eps)k(\eps')}, \nonumber
\end{equation}
which, as expected,  is a smooth function of  $\eps$ and $\eps'$. The second term in this expression was previously obtained in Ref.~\onlinecite{shuck} via a slightly different method which misses the first term of the right hand side of Eq.(\ref{I3D}.

A similar analysis in 2d is more subtle due to divergence of the integration in the direction perpendicular to the boundary. However, there is a natural upper limit for this integration given by the linear system size $L$.
Upon using $k(\eps)\simeq k_{F}(1+\eps/ 2\eps_F)$ and introducing the scaled perpendicular distance to the boundary $y=2k_{F}x$,
\begin{eqnarray}
\label{I2D}
\bar{I}^{{\rm short}}_{2d}(\eps,\eps')&&=\frac{{\cal L}}{2k_{F}{A}} \times \\
\int_{0}^{2k_{F}L}J_{0}\left[\left(1+\frac{\eps}{2 \eps_F}\right)y\right]
&&J_{0}\left[\left(1+\frac{\eps'}{2 \eps_F}\right)y\right]dy. \nonumber
\end{eqnarray}
Employing the asymptotic expression for the Bessel functions, we find
\begin{eqnarray}
\label{Ia}
&& \bar{I}^{{\rm short}}_{2d}(\eps,\eps') =\frac{{\cal L}}{2k_{F}{A}} \times \\
&& \left[C+\frac{1}{\pi}\int_{1}^{2k_{F}L}\frac{\sin{2y}+\cos{2(\eps-\eps')y/\eps_F}}{y}dy \right] \nonumber
\end{eqnarray}
valid for $k_{F}L \gg 1$. In Eq.~(\ref{Ia}) the constant $C=\int_{0}^{1}J_{0}^{2}(y)dy \simeq 0.850 \ldots$. We see that, contrary to the 3d case, $\bar{I}^{{\rm short}}_{2d}$  depends on energy (through the difference $\eps-\eps'$). This implies that in 2d chaotic systems the superconducting gap is energy dependent even to leading order in $\zeta$.

The integrals in Eq.~(\ref{Ia}) can be expressed in terms of the sine-integral (Si) and cosine-integral (Ci) functions. Our final result is
\begin{equation}
\label{ISIL}
\begin{cases}
\bar{I}^{{\rm short}}_{3d}(\eps_F)=\frac{\pi\cal{S}}{4k_{F}{V}} & {\rm 3d}, \\
 & \\
\bar{I}^{{ \rm short}}_{2d}(\eps_F,\eps - \eps')=\frac{\cal{L}}{k_{F}{A}}\left[C'+\frac{{\rm Si}(4k_{F}L)}{\pi}\right] & {\rm 2d} \\
 +\frac{{\cal L}}{2\pi k_{F}{A}}\left[{\rm Ci}\left(\frac{4(\eps-\eps')k_{F}L}{\eps_F}\right)-{\rm Ci}(\frac{2(\eps-\eps')}{\eps_F})\right]
\end{cases}
\end{equation}
with $C'=C-{\rm Si}(2)/ \pi=0.339 \ldots$. Thus for fixed $\eps$ and $\eps'$,  ${\bar I}^{{\rm short}}$ scales with $\zeta = 1/k_FL \ll 1$ as follows
\begin{equation}
\label{Iz}
\begin{cases}
\bar{I}^{{\rm short}}_{3d}(\eps_F)\propto \zeta+b\zeta^{2} & {\rm 3d}, \\
\bar{I}^{{ \rm short}}_{2d}(\eps_F,\eps - \eps')\propto \zeta+b'\zeta\log{\zeta} & {\rm 2d}, \\
\end{cases}
\end{equation}
where $b$ and  $b'$ are constants independent of the system size.
Note the non-algebraic dependence on $\zeta$ in the 2d case. The constant $b$ turns out to be much smaller than all other second order contributions to the gap, and will be dropped from now on.

Now we focus on the contribution of long paths, $\tilde{R}^{{\rm long}}$, to the spatial integral (\ref{Ismoo}). We use the expression for $\tilde{R}$ as a sum over classical closed paths $\gamma(\vec{r})$ starting and ending at $\vec{r}$ with length $L_{\gamma}(\vec{r})$. Now we impose the condition
\begin{equation}
L_{\gamma}(\vec{r})>>L,
\end{equation}
expressing the fact that the paths are long, namely, they hit the boundary several times. As is standard in these cases, we evaluate the smooth functions $D_{\gamma}$  in  $\tilde{R}^{{\rm long}}$ at the Fermi energy and expand  $k(\eps)\simeq k_{F}+k(\eps)^{2}/ 2k_{F}$ to get 
\begin{equation}
\label{DoubleSum}
\bar{I}^{{\rm long}}(\epsilon,\epsilon')=\Re\int\sum_{\gamma,\gamma'}^{l}D_{\gamma}D_{\gamma'}\frac{{\rm e \ }^{i\Phi_{\gamma,\gamma'}^{+}}+{\rm e \ }^{i\Phi_{\gamma,\gamma'}^{-}}}{4} d\vec{r}
\end{equation}
The phases involved in the spatial integration are (we do not include topological indexes for simplicity)
\begin{eqnarray}
\label{phas}
\Phi_{\gamma,\gamma'}^{\pm}(\eps,\eps',\vec{r})&=& k_{F}(L_{\gamma}\pm L_{\gamma'}) \\
&+&\frac{k_{F}}{2\eps_F}(L_{\gamma}\eps\pm L_{\gamma'}\eps') , \nonumber
\end{eqnarray}
where $L_{\gamma}=L_{\gamma}(\vec{r})$ is the length of the trajectory $\gamma$. 

In chaotic systems different trajectories, in general, will have lengths differing by at least $L$ (see, however \cite{loops}). This means that since the first term in Eq.~(\ref{phas}) scales as $1/\zeta \gg 1$, the integral over $r$ in  $\Phi_{\gamma,\gamma'}^{+}(\eps,\eps',\vec{r})$ and $\Phi_{\gamma,\gamma}^{\pm}(\eps,\eps',\vec{r})$ can be evaluated by the stationary phase method. Within this approximation, oscillatory integrals of the form
\begin{equation*}
\int f(x){\rm e \ }^{i \lambda h(x)}dx 
\end{equation*}
are given to leading order in $1/\lambda$ by
\begin{equation*}
\int f(x){\rm e \ }^{i \lambda h(x)}dx \simeq f(x^{*})\frac{{\rm e \ }^{i \lambda h(x^{*})}}{\sqrt{2 \pi \lambda h''(x^{*})}},
\end{equation*}
where $h'(x^{*})=0$, and therefore each spatial integration in Eq. (\ref{DoubleSum}) yields an extra factor $\propto 1/\zeta^{1/2}$. Combining this with the prefactors (\ref{Dz}), we find that the contribution of pairs  $\gamma \ne \gamma'$ (the so-called non-diagonal contribution) $\bar{I}^{{\rm long}}_{{\rm ndg}}$) is of order
\begin{equation}
\label{Indgz}
\bar{I}^{{\rm long}}_{{\rm ndg}}(\eps,\eps')\propto
\begin{cases}
\zeta^{5/2} & {\rm 3d}, \\
\zeta^{2} & {\rm 2d}.
\end{cases}
\end{equation}

On the other hand, terms that involve $\Phi_{\gamma,\gamma}^{-}(\eps,\eps',\vec{r})$ do not oscillate rapidly, because in this case the highly oscillatory terms in the phase cancel each other leaving the second term in Eq.~(\ref{phas}) which scales as $\zeta$ and not as $1/\zeta$.
\begin{equation}
\Phi_{\gamma,\gamma}^{-}(\eps,\eps',\vec{r})=\frac{k_{\rm F}L_{\gamma}(\vec{r})}{2\epsilon_{\rm F}}(\epsilon-\epsilon').
\end{equation}
This contribution involves coherent double sums over classical trajectories and is usually referred to as the diagonal contribution, $\bar{I}^{{\rm long}}_{{\rm dg}}$. Taking $\gamma = \gamma'$ in Eq.(\ref{DoubleSum}) we easily find
\begin{equation}
\label{IdgI}
\bar{I}^{{\rm long}}_{{\rm dg}}(\eps_F,\eps-\eps')=\int\sum_{\gamma}^{l}D_{\gamma}^{2}\cos{\Phi_{\gamma,\gamma}^{-}(\eps,\eps',\vec{r})}d\vec{r},
\end{equation}
which can be cast in a very compact form by introducing the purely classical function
\begin{equation}
\label{Pi}
\Pi_{l}(w)=\int\sum_{\gamma}^{l}D_{\gamma}^{2}\cos{wk_{F}L_{\gamma}(\vec{r})}d\vec{r},
\end{equation}
as follows
\begin{equation}
\label{Idg}
\bar{I}^{{\rm long}}_{{\rm dg}}(\eps_F,\eps-\eps')=
\begin{cases}
\frac{1}{{V}}\Pi_{l}\left(\frac{\eps-\eps'}{\eps_F}\right) & {\rm 3d} \\
\frac{1}{{A}}\Pi_{l}\left(\frac{\eps-\eps'}{\eps_F}\right) & {\rm 2d}.
\end{cases}
\end{equation}

Keeping also in mind the $\zeta$-dependence of the coefficients $D_{\gamma}$, we have
\begin{equation}
\label{Idgz}
\bar{I}^{{\rm long}}_{{\rm dg}}(\eps_F,\eps-\eps')\propto
\begin{cases}
\zeta^{2} & {\rm 3d}, \\
\zeta & {\rm 2d}.
\end{cases}
\end{equation}

Equations (\ref{ISIL},\ref{Indgz}) and (\ref{Idg}) complete the evaluation of ${\bar I}$. Restricting ourselves to the first two orders in $\zeta$ ($\zeta$  and $\zeta \log{\zeta}$ in the 2d case), we finally obtain
\begin{eqnarray}
\label{Ifin}
&&I(\eps,\eps')= \\
&&
\begin{cases}
\frac{\lambda}{{V}}\left[1+\bar{I}^{{\rm short}}_{3d}(\eps_F)-\frac{\pi^{2} {S}^{2}}{16k_{F}^{2}{V}^{2}}+\bar{I}^{{\rm long}}_{{\rm dg}}(\eps_F,\eps-\eps')\right]& {\rm 3d}, \\
& \\
\frac{\lambda}{{A}}\left[1+\bar{I}^{{\rm short}}_{2d}(\eps_F,\eps-\eps')+\bar{I}^{{\rm long}}_{{\rm dg}}(\eps_F,\eps-\eps')\right]& {\rm 2d}.
\end{cases}
\nonumber
\end{eqnarray}

Equations (\ref{Ifin}) together with the definitions (\ref{ISIL}) and (\ref{Idg}) allow for the calculation of interaction matrix elements in 3d and 2d chaotic grains. In general, the explicit evaluation of $\Pi_{l}(w)$  requires the precise knowledge of all classical paths up to lengths of the order of the single-particle coherence length $l$ that have a crossing at $\vec{r}$ for every point inside the cavity. However, if $l$ is large enough compared to $L$ (in practice $l \simeq 5L$ suffices)
ergodic arguments can be invoked and a closed expression for the interaction matrix elements can be found. In situations when $l \simeq L$ one must carry out the explicit system-dependent calculation.

Classical ergodicity of chaotic systems can be formulated in various ways \cite{almeida}, and we are going to give only a brief sketch of its consequences here. The main mechanism behind universality in the quantum mechanical description of classically chaotic systems, resides in the behavior of typical (in the sense of measure theory) classical trajectories. By definition, a typical trajectory of a chaotic system will explore in an uniform way the available phase space, thus implying the equivalence between temporal and microcanonical averages.

This uniformity extends, in a non-trivial way, to the periodic orbits as well. The key concept here is the classical probability of return, defined as
\begin{equation}
P(\vec{x}_{0},t,e)=\frac{1}{Z(e,t)}\delta(\vec{x}_{0}-\vec{x}(\vec{x}_{0},t))\delta(H(\vec{x}_{0})-e)
\end{equation}
where $\vec{x}_{0}=(\vec{r}_{0},\vec{p}_{0})$ is a point in phase space mapped at time $t$ into $\vec{x}(\vec{x}_{0},t)$ by the solution of the classical equations of motion. Clearly, the function $\delta(\vec{x}_{0}-\vec{x}(\vec{x}_{0},t))$ is non-zero only when the classical flow maps an initial point into itself after a time $t$ and plays the role of a probability of classical return. Moreover in case we want to select a fixed energy we use an extra condition given by the value of the Hamiltonian function along the trajectory. Finally, the probability must be normalized such that
\begin{equation}
\int P(\vec{x}_{0},t,e) d\vec{x}_{0}=1
\end{equation}
thus fixing $Z(t,e)$. The key observation here is that, by definition, the set of points where $P(\vec{x}_{0},t,e)$ is different from zero, belongs to periodic orbits with period $t$. Although the original ergodicity criteria were given in terms of typical trajectories, the theory of dynamical systems provides a strictly equivalent definition of ergodicity in terms of periodic orbits,
\begin{equation}
\label{Erg}
P(\vec{x}_{0},t,e) \to {\rm \ \ const. \ \ \ \ for \ } t \to \infty,
\end{equation}
That means that not only typical trajectories, but also periodic orbits uniformly fill the available phase space. We remark that the left hand side of this equation, a set of delta peaks at the periods of the classical periodic orbits, must be understood in the sense of distributions, namely, both sides are assumed to be integrated over a smooth function of time and phase-space position.

In order to make contact with the coordinate representation used so far, we use the uniformity of periodic orbits in phase space expressed by Eq. (\ref{Erg}), and integrate out the momentum. This integral can be exactly calculated \cite{martinp}. It involves a Jacobian of the form $\partial \vec{p}(t)\partial(\vec{r}_{0})$, which is indeed proportional to the semiclassical prefactors $D_{\gamma}$. In summary, in the present context classical ergodicity leads to the following {\it sum rule} \cite{martinp} for classical closed orbits,
\begin{equation}
\label{martin}
\sum_{\gamma}^{l \gg L}D_{\gamma}^{2}\delta(l-L_{\gamma}(\vec{r}))=
\begin{cases}
\frac{4 \pi^{2}}{k_{F}^{2}{V}} & {\rm 3d} \\
\frac{4}{k_{F}{A}} & {\rm 2d},
\end{cases}
\end{equation}
As was mentioned previously integration over lengths up to $l$ on both sides with a smooth weight function is also assumed.
Using this result and noting that the right hand side of Eq.~(\ref{martin}) is independent of the position $\vec{r}$, we  get
\begin{equation}
\Pi_{l \gg L}(w)=
\begin{cases}
\frac{4 \pi^{2}}{k_{F}^{3}}\frac{\sin{wk_{F}l}}{w} & {\rm 3d}, \\
\frac{4}{k_{F}^{2}}\frac{\sin{wk_{F}l}}{w} & {\rm 2d}.
\end{cases}
\end{equation}
In the ergodic regime, $l \gg L$, these results enable us to evaluate explicitly the energy dependence of the interaction matrix elements in chaotic cavities.




\begin{thebibliography}{99}
\bibitem{erg1} Roughly speaking, the quantum ergodicity theorem \cite{erg2} says that for classical systems with ergodic classical dynamics, the eigenfunctions converge to an uniformly distributed density in the semiclassical limit.
\end{thebibliography}
\end{document}